\documentclass[12pt]{article}
\usepackage{epsfig}
\usepackage{bezier}
\usepackage{bbm}
\usepackage{graphicx}
\newcommand{\lesssim}{\:\mbox{\raisebox{-3pt}{$\stackrel%
{\displaystyle <}{\sim}$}}\:}
\newcommand{\gtrsim}{\:\mbox{\raisebox{-3pt}{$\stackrel%
{\displaystyle >}{\sim}$}}\:}
\begin{document}
\title{\normalsize \hfill UWThPh-2003-16 \\[1cm] \LARGE
Neutrino Physics -- Theory\thanks{Lectures given at the
  41.\ Internationale Universit\"atswochen f\"ur Theoretische Physik,
\textit{Flavour Physics}, Schladming, Styria, Austria, February
22--28, 2003}}
\author{Walter Grimus \\
\small Institut f\"ur Theoretische Physik, Universit\"at Wien \\
\small Boltzmanngasse 5, A--1090 Wien, Austria \\
\small E-mail: walter.grimus@univie.ac.at}
\date{14 July 2003}
\maketitle
\begin{abstract}
We discuss recent developments in neutrino physics and 
focus, in particular, on neutrino oscillations and
matter effects of three light active neutrinos. Moreover, we discuss
the difference between Dirac and Majorana neutrinos, 
neutrinoless $\beta\beta$-decay, absolute neutrino masses and
electromagnetic moments. Basic mechanisms and a few models for
neutrino masses and mixing are also presented. 
\end{abstract}
\newpage
\tableofcontents
\newpage
\section{Motivation}
In recent years neutrino physics has gone through a spectacular 
development (for reviews see, for instance,
\cite{smirnov,goswami-review,giunti,drexlin}).  
Data concerning solar and atmospheric neutrino deficits
have been accumulated, these deficits have been established as neutrino
physics phenomena and the Solar Standard Model \cite{bahcall} has been
confirmed. The last steps of this exciting development were the
results of the SNO \cite{SNO} and KamLAND experiments \cite{KamLAND}:
The SNO experiment provided a model-independent 
proof of solar $\nu_e \to \nu_{\mu,\tau}$ transitions and the
terrestrial disappearance of $\bar\nu_e$ reactor neutrinos in the
KamLAND experiment has shown that the puzzle of the solar neutrino
deficit is solved by neutrino oscillations \cite{pontecorvo}. This
gives us confidence that the same is true for the 
atmospheric neutrino deficit as well. For information on the 
history of neutrino oscillations see \cite{bilenky,okun}, for the
recent experimental history see the contribution of G. Drexlin to
these proceedings \cite{drexlin}. General reviews on
neutrino physics can be found, for instance, in
Refs.~\cite{BP87,BGG99,reviews,books}. 

These lecture notes are motivated by this development and aim at
supplying the theoretical background for understanding and assessing
it. In view of the importance of neutrino oscillations in this
context, we will give a thorough discussion of vacuum 
neutrino oscillations and matter effects \cite{wolfenstein,MSW} 
(see Section \ref{neuosc});
the description of the latter is tailored for an understanding of the
flavour transformation of solar neutrinos and effects in earth matter.
Then, in Section \ref{DvsM}, we will switch to the subject of the neutrino
nature, which is a question of principal interest but has no impact on
neutrino oscillations; we will work out the difference between Dirac and 
Majorana neutrinos and the basics of Majorana neutrino 
effects. Eventually, in Section \ref{models}, we will come to the least
established field of neutrino physics: models for neutrino masses and
mixing. In view of the huge number of models and textures, we cannot
try to cover the field but rather discuss a small selection of basic
mechanisms for generating neutrino masses and mixing. This selection
will necessarily be biased due to personal interest and prejudices. 
A similar judgement has to be made concerning the literature quoted in
these lecture notes; owing to the host of papers which have appeared
in recent years only a small selection can be quoted here (for
literature on neutrino experiments see Ref.~\cite{drexlin}). 
Finally, we present conclusions in Section \ref{summary}.

Abbreviations used in these lecture notes: 
CC = charged current, NC = neutral current, 
LBL = long baseline, SBL = short baseline, 
MSW = Mikheyev-Smirnov-Wolfenstein, LMA = large mixing angle, 
MM = magnetic moment, EDM = electric dipole moment, 
SM = Standard Model, SUSY = supersymmetry,
GUT = Grand Unified Theory, 
VEV = vacuum expectation value.

\section{Neutrino oscillations}
\label{neuosc}
\subsection{Neutrino oscillations in vacuum}
\label{neuoscvac}
Here we give a simple and yet quite physical 
derivation of the neutrino oscillation formula.
The first observation is that---as the NC interactions
are flavour-blind---neutrino-flavour production and detection proceeds 
solely via CC interactions with the Hamiltonian density
\begin{equation}\label{Hcc}
\mathcal{H}_{cc} = \frac{g}{\sqrt{2}} W_\rho^- \sum_{\alpha = e,\mu,\tau}
\bar\ell_\alpha \gamma^\rho \nu_{\alpha L} + \mathrm{H.c.}
\end{equation}
The next observation is that flavour transitions are induced by 
neutrino mixing:
\begin{equation}\label{mixing}
\nu_{\alpha L} = \sum_j U_{\alpha j} \nu_{jL} \,.
\end{equation}
This relation means that the left-handed flavour fields are not
identical with the left-handed components of the neutrino mass
eigenfields $\nu_{j}$ corresponding 
to the mass $m_j$, but are related via a matrix $U$, which is
determined by the neutrino mass term which will be discussed in
Section 3. Here we simply assume the existence of the mixing matrix
$U$ and the neutrino masses $m_j$.

We confine ourselves, apart from a few side remarks, to the following 
basic assumptions:
\begin{itemize}
\item There are three active neutrino flavours; 
\item The mixing matrix $U$ is a $3 \times 3$ unitary matrix. 
\end{itemize}
These assumptions are supported by the results of the neutrino
oscillation experiments \cite{drexlin}. 

A crucial observation is that neutrino flavour is defined by the
associated charged lepton in production and detection processes; there
is no other physical way to define neutrino flavours. 
Looking at Eqs.~(\ref{Hcc}) and (\ref{mixing}), we find that, in a reasonable
approximation, neutrino-flavour states are given by
\begin{equation}\label{flavourstate}
| \nu_\alpha \rangle = \sum_{j=1}^3 U_{\alpha j}^* | m_j \rangle \,,
\end{equation}
where the mass eigenstates $| m_j \rangle$ fulfill the orthogonality
condition 
\begin{equation}
\langle m_ j | m_k \rangle = \delta_{jk} \,.
\end{equation}
The state (\ref{flavourstate}) describes a neutrino with flavour
$\alpha$ sitting at the position $x = 0$. 
Every mass eigenstate propagates as a plane wave. If we have a
stationary neutrino state with neutrino energy $E$, we derive from
Eq.~(\ref{flavourstate}) the 
following form of the propagating state:
\begin{equation}\label{propstate}
| \nu_\alpha, x \rangle = \sum_{j=1}^3 U_{\alpha j}^* 
e^{-i(Et - p_j x)} | m_j \rangle \,,
\end{equation}
with neutrino momenta
\begin{equation}\label{momentum}
p_j = \sqrt{E^2 - m_j^2} \simeq E - \frac{m_j^2}{2E} 
\quad \mathrm{for} \quad E \gg m_j \,.
\end{equation}
The latter inequality indicates the relativistic limit of massive
neutrinos which is the relevant limit for all neutrino oscillation
experiments. 

Since the state
$|\nu_\alpha, x \rangle$ has flavour $\alpha$ at $x=0$, 
we compute the probability to find the flavour $\beta$ at $x=L$ by 
$| \langle \nu_\beta | \nu_\alpha, L\rangle|^2$. With the explicit form
(\ref{propstate}) of $|\nu_\alpha, x \rangle$ and with Eq.~(\ref{momentum}) 
we readily derive the
standard formula of the neutrino transition or survival probability: 
\begin{equation}\label{P}
P_{\nu_\alpha \to \nu_\beta} (L/E) =
\Bigg| \sum_{j=1}^3 U_{\beta j} U_{\alpha j}^* e^{-im_j^2L/2E}
\,\Bigg|^2.
\end{equation}

It is easy to convince oneself that in the case of antineutrinos 
one simply has to make the replacement $U \to U^*$ in Eq.~(\ref{P}).

Looking at Eq.~(\ref{P}), we read off the following properties of
$P_{\nu_\alpha \to \nu_\beta}$: 
\begin{itemize}
\item It describes a violation of family lepton numbers;
\item It is a function of mass-squared differences, e.g. of 
$m_2^2 - m_1^2$ and $m_3^2 - m_1^2$;
\item It has an oscillatory behaviour in $L/E$;
\item The relation 
$P_{\nu_\alpha \to \nu_\beta} = P_{\bar\nu_\beta \to \bar\nu_\alpha}$
is fulfilled as a reflection of CPT invariance; 
\item $P_{\nu_\alpha \to \nu_\beta}$ is invariant under the transformation 
$U_{\alpha j} \to e^{i\phi_\alpha}\, U_{\alpha j}\, e^{i\phi_j}$
with arbitrary phases $\phi_\alpha$, $\phi_j$.
\end{itemize}

Specializing Eq.~(\ref{P}) to 
2-neutrino oscillations with flavours $\alpha \neq \beta$, one
obtains the particularly simple formulas
\begin{eqnarray}
&&
U = \left( \begin{array}{rr} \cos \theta & \sin \theta \\ 
                            -\sin \theta & \cos \theta
           \end{array} \right), \label{U2} \\
&&
P_{\nu_\alpha \to \nu_\beta} = 
\sin^2 2\theta \times \frac{1}{2} \left(
1 - \cos \frac{\Delta m^2 L}{2E} \right), \label{2tran} \\
&&
P_{\nu_\alpha \to \nu_\beta} = P_{\nu_\beta \to \nu_\alpha}\,, \\
&&
P_{\nu_\alpha \to \nu_\alpha} =
P_{\nu_\beta \to \nu_\beta} =
1 - P_{\nu_\alpha \to \nu_\beta} \,. \label{2surv}
\end{eqnarray}
In Eq.~(\ref{U2}) the rephasing invariance of 
$P_{\nu_\alpha \to \nu_\beta}$ has been used to reduce $U$ to a
rotation matrix. 
For 2-neutrino oscillations the probabilities for neutrinos and
antineutrinos are the same.

The oscillation phases in Eq.~(\ref{P}) have the form 
\begin{equation}\label{oscphase}
\frac{\Delta m^2 L}{2E} = 2.53 \times 
\left( \frac{\Delta m^2}{1\, \mathrm{eV}^2} \right) \times
\left( \frac{1\, \mathrm{MeV}}{E} \right) \times
\left( \frac{L}{1\, \mathrm{m}} \right),
\end{equation}
where $\Delta m^2$ is one of the mass-squared differences.
Alternatively, the oscillation phase is expressed by the oscillation
length $L_\mathrm{osc}$ through
\begin{equation}\label{Losc}
\frac{\Delta m^2 L}{2E} \equiv 2\pi\, \frac{L}{L_\mathrm{osc}} 
\;\; \mathrm{with} \;\;
L_\mathrm{osc} = 2.48\, \mathrm{m}\, \times
\left( \frac{E}{1\, \mathrm{MeV}} \right) \times
\left( \frac{1\, \mathrm{eV}^2}{\Delta m^2} \right). 
\end{equation}

In a neutrino oscillation experiment, the sensitivity to $\Delta m^2$
is determined by the requirement that the phase (\ref{oscphase}) is of
order one or not much smaller; this requirement depends on
the characteristic ratio $L/E$ relevant for the experimental setup. 
By definition, SBL experiments are characterized by a sensitivity to
$\Delta m^2 \gtrsim 0.1$ eV$^2$, while terrestrial experiments with 
a sensitivity below 0.1 eV$^2$ are called LBL experiments. 
Note that using longer baselines or smaller energies 
moves the sensitivity to smaller mass-squared
differences (see Eq.~(\ref{oscphase})).

No neutrino oscillation experiment
has found a mass-squared difference $\Delta m^2 \gtrsim 0.1$ eV$^2$,
apart from the LSND experiment \cite{LSND}. 
This experiment, however, has not not been confirmed by any other 
experiment and requires a forth neutrino which has to be sterile
because the invisible Z width at LEP accommodates exactly the three
active neutrinos. A \emph{sterile neutrino} is defined as a neutral
massive fermion which mixes with the active neutrinos, but has
negligible couplings to the W and Z bosons. The result of the LSND
experiment is at odds \cite{schwetz} with the combined data sets of
either all other 
SBL experiments or the solar and atmospheric neutrino experiments,
when interpreted in terms of 4-neutrino oscillations. See, however,
Ref.~\cite{conrad}, which shows that a 5-neutrino interpretation 
might reconcile the LSND result with the negative results of the other
SBL experiments. 

Let us discuss some concrete examples for the sensitivity to $\Delta m^2$.
For reactor neutrinos the energy is about $E \sim 1$ MeV. Therefore, 
with $L \sim 1$ km, the sensitivity is 
$\Delta m^2 \sim 10^{-3}$ eV$^2$; this is the case of the CHOOZ
\cite{CHOOZ} and Palo Verde \cite{paloverde} experiments. 
On the other hand, the KamLAND reactor experiment \cite{KamLAND} with
$L \sim 100$ km is sensitive down to
$\Delta m^2 \sim 10^{-5}$ eV$^2$. Experiments exploiting the source of 
atmospheric  neutrinos with typical energies $E \sim 1$ GeV have as
maximal baseline the diameter of the earth, i.e. $L \lesssim 10^4$
km, from where the sensitivity 
$\Delta m^2 \gtrsim 10^{-4}$ eV$^2$ follows. 
Solar neutrinos have a very long baseline of $L \simeq 150 \times
10^6$ km and rather low energies of $E \sim 1 \div 10$ MeV. Thus, in
principle with solar neutrino experiments one can reach 
$\Delta m^2 \sim 10^{-11}$ eV$^2$.

\paragraph*{Quantum-mechanical aspects of neutrino oscillations:}
Our phenomenological de\-rivation of the oscillation probability
(\ref{P}) needs some quantum-mechanical support 
(for extensive reviews see Refs.~\cite{zralek,beuthe1}).
The points
of our derivation which need justification are the following:
\begin{itemize}
\item
We have employed stationary neutrino states, i.e., the plane waves
have the same energy $E$ but different momenta $p_j$;
\item
The usage of plane waves for the neutrino mass eigenstates raises the
question how this is compatible with the localization of
neutrino source and detection in space.
\end{itemize}
In order to find limitations of the canonical oscillation probability
we have to use a mathematical picture which is as close as possible to
the actual situation of an experiment. Neutrinos are never directly
observed; thus, for the description of neutrino oscillations it is
required to consider the complete neutrino 
production--detection chain and to use only those quantities or particles
which are really observed or manipulated in an experiment \cite{rich}.
Consequently, we are lead to consider neutrino production and detection as a
single big Feynman diagram where both the source and detector particle are
described by wave packets which are localized in space, whereas the
neutrinos are presented by inner lines of the big diagram; such models are
called \emph{external wave packet models} \cite{beuthe1,beuthe2} and
the big diagram is treated in a field-theoretical way
\cite{GKLL}. However, due to the macroscopic distance between source
and detector the neutrinos are described as on-shell although they are
on an inner line of the Feynman diagram \cite{stocki1}.

We start with a consideration of the detector particle which is
central to the first issue above \cite{lipkin1,stocki2,lipkin2}. In
all experiments performed so far the detector particle is a bound
state with energy $E_D$, either a nucleon in a nucleus or an electron in
an atom. Such states are energy eigenstates. In addition, the detector
bound state will experience thermal random movements with a whole spectrum of
energies related to the temperature. However, there are no phase
correlations between the different energy components of the thermal
energy distribution which is, therefore, very well described by a density
matrix which is diagonal in energy and the summation over this energy
distribution is performed in the \emph{cross section} corresponding to
the big production--detection Feynman diagram
\cite{lipkin2,stodolsky}. In other words, in the \emph{amplitude}
corresponding to the big diagram, we have a definite detector particle
energy $E_D + \varepsilon^D_\mathrm{thermal}$, with an
\emph{incoherent} summation over 
the distribution of $\varepsilon^D_\mathrm{thermal}$ in the cross section.
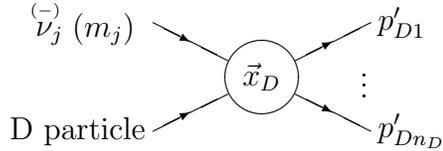
\begin{figure}[t]
\begin{center}
\setlength{\unitlength}{5mm}
\begin{picture}(12,4)(5,2)%(3,2)
\put(11,4){\circle{2}} \put(11,4){\makebox(0,0){$\vec{x}_D$}}
\put(11.8944,4.4472){\vector(2,1){1}}
\put(12.8944,4.9472){\line(2,1){1}}
\put(11.8944,3.5528){\vector(2,-1){1}}
\put(12.8944,3.0528){\line(2,-1){1}}
\put(8.1056,2.5528){\vector(2,1){1}}
\put(9.1056,3.0528){\line(2,1){1}}
% neutrino line
\put(8.1056,5.4472){\vector(2,-1){1}}
\put(9.1056,4.9472){\line(2,-1){1}}
%
%%%%%  inscriptions
% neutrino line
\put(7.9056,5.4472){\makebox(0,0)[r]{
$\stackrel{\scriptscriptstyle (-)}{\nu}_{\hskip-3pt j}(m_j)$
}}
% D
\put(7.9056,2.5528){\makebox(0,0)[r]{D particle}}
\put(14.0944,5.4472){\makebox(0,0)[l]{$p'_{D1}$}}
\put(14.0944,2.5528){\makebox(0,0)[l]{$p'_{D{n_D}}$}}
\put(13.6,4){\makebox(0,0)[l]{$\vdots$}}
\end{picture}
\end{center}
\caption{The particles participating in the detector reaction at 
$\vec x_D$. \label{detector}} 
\end{figure}
Concerning the $n_D$ final states in the detector reaction 
(see Fig.~\ref{detector}), 
we note that energy/momentum measurements are performed and that it is
again summed incoherently over these energies $E'_{Dk}$ in the cross
section. Thus, looking at Fig.~\ref{detector} we come to the conclusion
that in the \emph{amplitude} the neutrino energy $E$ is fixed and
given by 
\begin{equation}
E_\nu = \sum_k E'_{Dk} - E_D -
  \varepsilon^D_\mathrm{thermal} \,.
\end{equation}
Since the oscillation probability (\ref{P})  is directly derived from
the cross section, we come to following conclusion which has, in
particular, been stressed by H.\ Lipkin 
\cite{stocki1,lipkin1,stocki2,lipkin2,stodolsky}: \\
\begin{center}
\fbox{\parbox{8.5cm}{Neutrinos with the \emph{same energy} $E$ but
different momenta $p_j = \sqrt{E^2 - m_j^2}$ are coherent.}}
\end{center} 
Moreover, summation over the neutrino energies $E$ is effectively 
performed in the
cross section, i.e., it is an incoherent summation and no wave packets
are associated with neutrinos of definite mass.

Now we address the justification of the second point. 
Whereas it is necessary to assume localization in space of the neutrino source
and detector wave functions, 
we stress that the final states in the source
and detection reaction can be taken as plane waves; 
the reason is, as mentioned above, that
the measurements performed with them are energy/momentum measurements
and for calculating the cross section corresponding to the big Feynman
diagram one has to sum \emph{incoherently}, i.e. in the cross section, 
over the regions in phase
space subject to kinematical restrictions according to the experiment.
Denoting the detector particle wave function in phase space by 
$\tilde \psi_D(\vec p\,)$, a reasonable range for the width of this
function is given by $\sigma_D \sim 10^{-3} \div 100$ MeV, where these
limits are given for a detector particle bound in an atom or in a
nucleus, respectively.
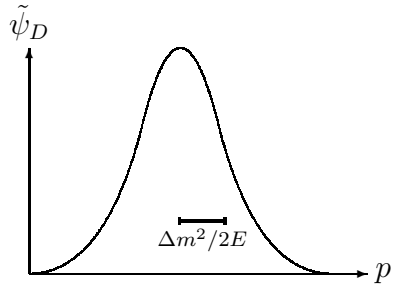
\begin{figure}[t]
\begin{center}
\setlength{\unitlength}{1cm}
\begin{picture}(4,4)
\put(0,0){\vector(1,0){4.5}}
\put(0,0){\vector(0,1){3}}
\bezier{200}(1.5,2)(2,4)(2.5,2)
\bezier{200}(2.5,2)(3,0)(4,0)
\bezier{200}(0,0)(1,0)(1.5,2)
\put(0,3.1){\makebox(0,0)[b]{$\tilde \psi_D$}}
\put(4.6,0){\makebox(0,0)[l]{$p$}}
\thicklines
\put(2,0.7){\line(1,0){0.6}}
\put(2,0.65){\line(0,1){0.1}}
\put(2.6,0.65){\line(0,1){0.1}}
\put(2.3,0.6){\makebox(0,0)[t]{$\scriptstyle \Delta m^2/2E$}}
\end{picture}
\end{center}
\caption{The wave function of the detector particle. The vertical line
  indicates the momentum difference $|p_2 - p_1|$. \label{wavefunction}}
\end{figure}
In Fig.~\ref{wavefunction} we have symbolically drawn a detector
particle wave function. Looking at this 1-dimensional 
picture it is easy to find the
condition for the existence of oscillations: Assuming for simplicity only
two neutrinos, the neutrino mass eigenstates have the same energy $E$ but
different momenta $p_1$, $p_2$ (\ref{momentum}). 
Suppose that we fix the momenta of the
final states of the detector process by $p'_{Dk}$ ($k = 1, \ldots,
n_D$, see Fig.~\ref{detector}). By momentum conservation, the 
values of the detector wave function relevant for the amplitude of the
complete production--detection cross section are
given by $\tilde \psi_D \left(\sum_k \vec{p}\,'_{Dk} - p_j \vec
\ell\,\right)$ for each 
neutrino mass eigenstate, where $\vec \ell$ is a unit vector pointing
from the source to the detector particle. Let us assume now that 
$\sum_k \vec{p}\,'_{Dk} - p_1 \vec \ell \simeq 0$. In order to obtain
coherence between the neutrino states with mass $m_1$ and $m_2$, the
momentum $p_2$ must \emph{not} fulfill $|p_2 - p_1| \gg \sigma_D$, because in
that case we have $\tilde \psi_D \left(\sum_k \vec{p}\,'_{Dk} - p_2
\vec \ell\,\right) \simeq 0$, the two 
mass eigenstates do not interfere and there are no neutrino
oscillations. Therefore, we are lead to the condition 
\begin{equation}\label{momcond}
|p_2 - p_1| \simeq \Delta m^2/2E \lesssim \sigma_D
\end{equation}
for neutrino oscillations---see Fig.~\ref{wavefunction}.
Denoting the width of the detector wave function in coordinate
space by $\sigma_{xD}$, then with Heisenberg's uncertainty relation
$\sigma_{xD} \sim 1/2\sigma_D$ the condition (\ref{momcond}) is 
rewritten as \cite{stocki1,stocki2,kayser} 
\begin{equation}
\sigma_{xD} \lesssim L_\mathrm{osc}/(4\pi) \,.
\end{equation}
This condition is trivially fulfilled because 
$\sigma_{xD}$ is microscopic whereas $L_\mathrm{osc}$ is macroscopic!
Similar considerations can be made for the neutrino source, with an
additional condition taking into account that the source must be
unstable \cite{stocki1,stocki2,stocki3}. 
Thus we can have confidence that the oscillation probability (\ref{P})
holds for practical purposes.

The present discussion is certainly not the most general one but, as
we believe, it is reasonably close to practical applications. It does not
include the discussion of a coherence length, i.e. a maximal distance
between source and detector until which coherence between different
mass eigenstates is maintained. There is a vast literature on that
point, see for instance
Refs.~\cite{beuthe1,beuthe2,GKLL,nussinov,giunti03} and citations therein.
One detail can, however, be immediately deduced from our discussion: The
better one determines experimentally the energies of all final states
in the detection reaction, the better one determines the neutrino
energy $E$. Since the coherence length is approximately obtained by 
$L_\mathrm{osc} \bar E/\Delta E$ \cite{nussinov}, where $\bar E$ is
the average neutrino energy and $\Delta E$ the neutrino energy spread,
it is evident that by pure detector manipulations one can influence
the coherence length and, in principle, by infinitely precise
measurements of the energies of the final particles in the detector
process, one could make it arbitrarily long \cite{stocki2,nussinov}.

Some steps in the above investigation of the validity of the oscillation
probability (\ref{P}) have been criticized in the literature---see, for
instance, Refs. \cite{beuthe1,beuthe2,giunti03}---and even the external
wave packet model is not undisputed \cite{giunti03}. This suggests
further investigations into the validity conditions of
Eq.~(\ref{P}). However, since all  
attempts up to now to find limitations of (\ref{P}) accessible to
experiment were in vain, the standard formula for the probability of neutrino
oscillations in vacuum seems to be very robust.

\subsection{Matter effects}

Matter effects \cite{wolfenstein,MSW,MS-review} 
play a very important role in neutrino oscillations. 
Here we confine ourselves to the case of 
ordinary matter, which is non-relativistic, electrically neutral and
without preferred spin orientation. It 
consists of electrons, protons and neutrons, i.e. fermions 
$f = e^-$, $p$, $n$. We denote the corresponding 
matter (number) densities by $N_e = N_p$ and $N_n$.

We want to present a simple heuristic and straightforward derivation
of matter effects. First we notice that the vacuum oscillation
probability (\ref{P}) can be derived by using the Hamiltonian 
\begin{equation}\label{Hvac}
H_\nu^\mathrm{vac} = \frac{1}{2E} U \hat{m}^2 U^\dagger \,,
\end{equation}
where $\hat{m}$ is the diagonal matrix of neutrino masses. In order to
obtain the effective Hamiltonian in matter, we have to add to
$H_\nu^\mathrm{vac}$ a term describing matter effects.
With the above properties of ordinary matter we have the expectation
values 
\begin{equation}\label{expectation}
\langle \bar f_L \gamma_\mu f_L \rangle_{\mathrm{matter}} = 
\frac{1}{2} N_f \delta_{\mu 0} \,.
\end{equation}
Adding CC and NC contributions for neutrinos in the background of
ordinary matter, the SM of electroweak interactions together with 
Eq.~(\ref{expectation}) provides us with the Hamiltonian density
\begin{eqnarray}
\mathcal{H}^{\mathrm{mat}} & = & 
\frac{G_F}{\sqrt{2}} \sum_{\alpha=e,\mu,\tau} \nu_\alpha^\dagger
(\mathbbm{1} - \gamma_5) \nu_\alpha \nonumber \\
&& \times \sum_f N_f(\delta_{\alpha f} + T_{3f_L} -
2 \sin^2 \theta_W Q_f) \,,
\end{eqnarray}
where $\delta_{\alpha f}$ comes from a Fierz transformation of the CC
interaction,\footnote{This term is non-zero only for $\nu_e$
  interacting with the background electrons.} $\theta_W$ is the weak
mixing angle, and $T_{3f_L}$ and $Q_f$ are the weak isospin and
electric charge of the fermion $f$, respectively. Thus we arrive at
the effective flavour Hamiltonian \cite{wolfenstein,MSW}
\begin{equation}\label{Heff}
H^\mathrm{mat}_\nu = 
\frac{1}{2E} U \hat{m}^2 U^\dagger + 
\sqrt{2}\, G_F\, \mathrm{diag}\, 
\left( N_e - \frac{1}{2} N_n, - \frac{1}{2} N_n, - \frac{1}{2} N_n
\right).
\end{equation}
Then the evolution of a neutrino state undergoing oscillations and
matter effects is calculated by solving 
\begin{equation}\label{evolution}
i \frac{da}{dx} = H_\nu^\mathrm{mat}\, a \quad \mathrm{with} \quad
a = \left( \begin{array}{c} a_e \\[-1mm] a_\mu \\[-1mm] 
a_\tau \end{array} \right),
\end{equation}
where $a_e$, $a_\mu$ and $a_\tau$ are the amplitudes of electron, muon
and tau neutrino flavours.

For active neutrinos ($\nu_\alpha$ with $\alpha = e, \mu, \tau$) the
neutron density $N_n$ can be dropped in the effective Hamiltonian,
since it does not induce flavour transitions. Thus for active
neutrinos only the electron density is relevant and there are no
matter effects in 2-neutrino $\nu_\mu \to \nu_\tau$ transitions.
If sterile neutrinos exist, $N_n$ does affect $\nu_\alpha \to \nu_s$
because sterile neutrinos do not experience matter effects, i.e.,
$N(\nu_s)=0$ in $H^\mathrm{mat}_\nu$. 

If we consider 2-neutrino oscillations, we obtain from
Eq.~(\ref{Heff}) the well-known 2-flavour Hamiltonian 
by subtraction of an irrelevant diagonal matrix and by using the 
$2 \times 2$ mixing matrix (\ref{U2}) \cite{bethe}:
\begin{equation}\label{H2}
H_\nu^\mathrm{mat} = \frac{1}{4E}
\left( \begin{array}{cc} 
A - \Delta m^2 \cos 2\theta & \Delta m^2 \sin 2\theta \\
\Delta m^2 \sin 2\theta & -A + \Delta m^2 \cos 2\theta 
\end{array} \right) 
\end{equation}
with
\begin{equation}\label{A}
A = 2\, \sqrt{2}\, G_F E \left[ N(\nu_\alpha)-N(\nu_\beta) \right] 
\end{equation}
and $N(\nu_e) = N_e - N_n/2$, $N(\nu_\mu) = N(\nu_\tau) = -N_n/2$, 
$N(\nu_s) = 0$.
This equation shows once more what we have discussed above. 
From now on we do not discuss 
sterile neutrinos anymore in these lecture notes.

Performing all the analogous procedures for antineutrinos, we obtain 
$H^\mathrm{mat}_{\bar\nu}$ by making  
the replacements $U \to U^*$, matter term $\to$ $-$matter term
in $H^\mathrm{mat}_\nu$ of Eq.~(\ref{Heff}).

In order to have an idea of the strength of matter effects, we 
estimate the ``matter potential'' $A$ for the two most important
cases, the sun and the earth:
\begin{equation}\label{strength}
A = 2\sqrt{2} G_F E N_e \simeq 
\left\{ \begin{array}{l} 
\mbox{core of the sun:} \\
1.5 \times 10^{-5} \: \mbox{eV}^2 \,
\left( \frac{E}{1 \; \mathrm{MeV}} \right), \\
\mbox{earth matter:} \\
2.3 \times 10^{-7} \: \mbox{eV}^2 \: 
\left( \frac{\rho}{3\, \mathrm{g}\, \mathrm{cm}^{-3}} \right)
\left( \frac{E}{1 \; \mathrm{MeV}} \right).
        \end{array} \right.
\end{equation}
The matter potential is the quantity which has to be compared with
$\Delta m^2$ in order to estimate the influence of matter on neutrino
oscillations. 

In the evolution equation (\ref{evolution}) the Hamiltonian is in
general a function of the space coordinates through the electron
density $N_e(x)$. Looking at the 2-neutrino Hamiltonian (\ref{H2}), a 
MSW resonance \cite{MSW} occurs 
if there is a coordinate $x_\mathrm{res}$ such that 
\begin{equation}\label{res}
A(x_\mathrm{res}) = \Delta m^2 \cos 2\theta
\end{equation}
holds. In this case, the probability of flavour transitions can be
very big, in particular, if the neutrino goes adiabatically through
the resonance point, as we shall see in the next subsection. 

In the following, we will adopt---without loss of generality---the
convention $\Delta m^2 >0$ and $0^\circ \leq \theta \leq 90^\circ$
in the Hamiltonian (\ref{H2}). 
Therefore, the resonance condition is fulfilled for 
$\theta < 45^\circ$ and a suitable electron density. 

\paragraph*{On the validity of the effective flavour Hamiltonian:}
The heuristic derivation of matter effects resulting in the
Hamiltonian (\ref{Heff}) did not allow to see under which conditions
Eqs.~(\ref{Heff}) and (\ref{evolution}) hold. 
We now want to discuss a method which gives us some insight into the
limitations of these equations. 
We follow Refs.~\cite{sakurai,liu}.
Suppose we have an incident wave propagating in $z$-direction which
traverses a slab of matter of ``infinitesimal'' thickness $d$. 
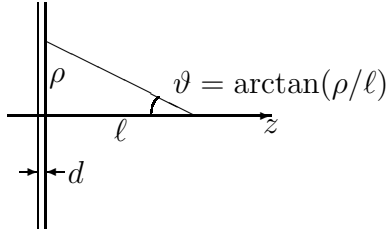
\begin{figure}[t]
\begin{center}
\setlength{\unitlength}{1cm}
\begin{picture}(4,3)
\put(-0.1,0){\line(0,1){3}}
\put(0,0){\line(0,1){3}}
\put(-0.5,1.5){\vector(1,0){3.5}}
\put(0,2.5){\line(2,-1){2}}
\put(0.05,2){\makebox(0,0)[l]{$\rho$}}
\put(1,1.45){\makebox(0,0)[t]{$\ell$}}
\put(-0.35,0.75){\vector(1,0){0.25}}
\put(0.25,0.75){\vector(-1,0){0.25}}
\put(0.3,0.75){\makebox(0,0)[l]{$d$}}
\put(3,1.45){\makebox(0,0)[t]{$z$}}
\bezier{40}(1.4,1.5)(1.4,1.68)(1.5,1.75)
\put(1.7,1.7){\makebox(0,0)[bl]{$\vartheta = \arctan (\rho/\ell)$}}
\end{picture}
\end{center}
\caption{Scattering at a slab of matter of thickness $d$ of an
  incident wave propagating in $z$-direction \label{scattering}}
\end{figure}
Then the incident wave undergoes multiple
scattering at the scatterers $f$ with number density $N_f$ in the
slab. According to Fig.~\ref{scattering} we 
sum over all scattered waves hitting a point on the $z$-axis at a distance
$\ell$ from the slab and obtain 
\begin{equation}\label{multiple}
\begin{array}{l} 
\displaystyle
N_f d \int_0^\infty 
\frac{e^{iE \sqrt{\rho^2 + \ell^2}}}{\sqrt{\rho^2 + \ell^2}}
\mathcal{A}_{\nu_\alpha f}(E,\vartheta)\, 2\pi \rho\, \mathrm{d}\rho = 
\\[4mm]
\displaystyle
2\pi N_f d \Bigg\{ 
\frac{e^{iE \sqrt{\rho^2 + \ell^2}}}{iE} \, 
\mathcal{A}_{\nu_\alpha f}(E,\vartheta) \Big|_0^\infty
- \\[4mm] 
\displaystyle 
\int_0^\infty \! \mathrm{d}\rho \,
\frac{e^{iE \sqrt{\rho^2 + \ell^2}}}{iE \ell\, [1 + (\ell/\rho)^2]} \, 
\frac{\partial \mathcal{A}_{\nu_\alpha f}(E,\vartheta)}{\partial \vartheta}
\Bigg\}.
\end{array}
\end{equation}
In this equation, $\mathcal{A}_{\nu_\alpha f}$ is the scattering
amplitude. Eq.~(\ref{multiple}) can be considerably
simplified. Firstly, it is
reasonable to assume that in the term in the
second line of this equation we can drop the
limit $\rho \to \infty$ because that contribution averages out.
Secondly, for 
$1/(E \ell) \ll 1$ and a smooth behaviour of the scattering amplitude,
we can also drop the term in the third line of
Eq.~(\ref{multiple}). Now we go one step further and consider a slab of finite 
thickness $D = kd$, where $d$ is ``infinitesimal'' whereas $k$ is a
large integer. With the above-mentioned approximations we compute the 
phase of incident + transmitted wave through the finite slab as 
\begin{equation}\label{finiteslab}
\begin{array}{l}
\lim_{k \to \infty} \left\{
1 + \frac{\displaystyle iED}{\displaystyle k} + 
\frac{\displaystyle 2\pi i N_f D 
\mathcal{A}_{\nu_\alpha f}(E,\vartheta=0)}{\displaystyle Ek} 
\right\}^k = 
\\[4mm]
\exp \left\{ i ED \left[1 + 2\pi N_f  
\mathcal{A}_{\nu_\alpha f}(E,\vartheta=0)/E^2 \right] \right\}.
\end{array}
\end{equation}
The expression in the curly brackets of the first line of this
equation, where the incident wave is represented by
$1 + iED/k \simeq \exp (iED/k)$,
corresponds to the ``infinitesimal'' slab of thickness 
$d = D/k$ and $\ell = 0$ (see Fig.~\ref{scattering}); 
the power $k$ indicates that the wave
propagates through $k$ such ``infinitesimal'' slabs. 
This consideration yields the well-known ``index of refraction''
\begin{equation}\label{index}
n(\nu_\alpha) = 1 + \frac{2\pi}{E^2} \sum_f N_f
\mathcal{A}_{\nu_\alpha f}(E,\vartheta=0) \,,
\end{equation}
where only the forward-scattering amplitude, i.e. $\vartheta = 0$
contributes. 
In $n(\nu_\alpha)$ we have indicated that different types of
scatterers can be distributed in the slab. Furthermore, the derivation
of the ``index of refraction'' is evidently not specific to neutrino
scattering. 

\begin{table}[t]
\begin{center}
\setlength{\tabcolsep}{3mm}
\renewcommand{\arraystretch}{1.2}
\begin{tabular}{|c|ccc|}\hline
$g_V$            & $e^-$ & $p$ & $n$ \\ \hline
$\nu_e$          & $2\sin^2 \theta_w + 1/2$  & 
 $-2\sin^2 \theta_w + 1/2$ & $-1/2$ \\ 
$\nu_{\mu,\tau}$ & $2\sin^2 \theta_w - 1/2$  
& $-2\sin^2 \theta_w + 1/2$ &
$-1/2$
\\ \hline
\end{tabular}
\end{center}
\caption{Vector coupling constants of the Hamiltonian density
  (\ref{Hdens}). \label{couplings}}
\end{table}
The connection with neutrino scattering is made by the Hamiltonian
density 
\begin{equation}\label{Hdens}
\mathcal{H}_{\nu_\alpha f} = \frac{G_F}{\sqrt{2}} \,
\bar \nu_\alpha \gamma_\rho (\mathbbm{1} - \gamma_5) \nu_\alpha \,
\bar f \gamma^\rho \left( g_V^{(\alpha,f)} - g_A^{(\alpha,f)} \gamma_5
\right) f \,.
\end{equation}
The coupling constants $g_V^{(\alpha,f)}$ are found in
Table~\ref{couplings}. The axial-vector coupling constants do not
contribute for ordinary matter (see Eq.~(\ref{expectation})). 
As mentioned before, 
in the $\nu_e e^-$ scattering amplitude, there are CC + NC contributions,
otherwise only NC interactions contribute. 
With the weak forward-scattering amplitude \cite{liu}
\begin{equation}\label{forward}
\mathcal{A}_{\nu_\alpha f}(E,\vartheta=0) = 
\frac{G_F E}{\sqrt{2}\pi}\, g_V^{(\alpha,f)} \,,
\end{equation}
Eq.~(\ref{finiteslab}) and Table~\ref{couplings}, we compute 
$2\pi \sum_f N_f \mathcal{A}_{\nu_\alpha f}(E,\vartheta=0)/E$ and obtain
exactly the matter terms as in 
$\mathcal{H}_\nu^\mathrm{mat}$ in Eq.~(\ref{Heff}).

Now we use this second derivation of the matter potential as a means
to check the validity of the Hamiltonian (\ref{Heff}) \cite{liu}.
We define two lengths: 
$d_\mathrm{scatt}$ is the average distance of scatterers in
ordinary matter, 
$d_\mathrm{var} \sim N_f(x)/\big| \frac{dN_f(x)}{dx} \big|$ is the
typical distance of matter density variations. Then we note the
following conditions for the validity of Eqs.~(\ref{Heff}) and
(\ref{evolution}):  
\begin{itemize}
\item[*] $E d_\mathrm{scatt} \gg 1 \; \Rightarrow \; 
\lambda_\nu^\mathrm{de\, Broglie} \ll 2\pi d_\mathrm{scatt}$ \,;
\item[*] $d_\mathrm{scatt} \ll d_\mathrm{var}$ \,;
\item[*] $d_\mathrm{scatt} \ll L_\mathrm{osc}$ \,.
\end{itemize}
The 2nd and 3rd condition arise from the requirement of describing the
neutrino state evolution by the differential equation
(\ref{evolution}). They are trivially fulfilled because 
$d_\mathrm{scatt}$ is microscopic, whereas $d_\mathrm{var}$ and $
L_\mathrm{osc}$ are macroscopic. For instance for the sun we
have $d_\mathrm{var} \sim 10^4$ km. 

The first condition arises from the dropping of the term in the
third line of Eq.~(\ref{multiple}). The quantity
$\lambda_\nu^\mathrm{de\, Broglie} = 2\pi/E$ 
is the de Broglie wave length of the neutrino.
First we have to estimate 
$d_\mathrm{scatt}$. It is easy the convince oneself that to a very good
approximation the electron density in matter is given by 
\begin{equation}\label{Ye}
N_e \simeq Y_e N_A (\rho/ 1\, g) \,,
\end{equation}
where $Y_e$ is the number of electrons per nucleon, 
$\rho$ is the matter density in units of g/cm$^3$ and 
$N_A \simeq 6.022 \times 10^{23}$ is Avogadro's constant.
Let us check the first condition for the sun. 
According to the Solar Standard Model \cite{bahcall}, in the solar
core one has $\rho \simeq 150$ and $Y_e \simeq 2/3$, from where it
follows that 
$N_e \simeq 100\, N_A$ cm$^{-3}$ and 
$d_\mathrm{scatt} = N_e^{-1/3} \simeq 0.25$ {\AA}. 
Thus the first
condition above is reformulated as 
$E\, d_\mathrm{scatt} \simeq 130\, E/(1\,\mathrm{MeV}) \gg 1$. 
For solar neutrinos with $E \gtrsim 1$ MeV this condition is
fulfilled. For earth matter this condition holds as well, because
$d_\mathrm{scatt}$ is a little larger than in the solar core. 

For other derivations of matter effects see Ref.~\cite{scharnagl} and
references therein.

\paragraph*{Survival probability for solar neutrinos:}
We want to conclude this subsection by some general
considerations concerning solar neutrinos within the framework of
2-neutrino oscillations. 
A neutrino produced in the solar core traverses first solar matter,
then travels through vacuum to the neutrino detector on earth; during
the night, the neutrino has to traverse, in addition, some stretch of
earth matter. We denote by 
$P^S_{e1,2}$ the probability for 
$|\nu_e \rangle \to | \nu_{1,2} \rangle$ in the sun. 
Furthermore, the probability of 
$|\nu_{1,2} \rangle \to | \nu_e \rangle$ transitions in earth matter
is called $P^E_{1,2e}$. Note that with the $2 \times 2$ mixing matrix
$U$ of Eq.~(\ref{U2}) during the day we have
\begin{equation}\label{day} 
P^E_{1e} = \cos^2 \theta
\quad \mathrm{and} \quad 
P^E_{2e} = \sin^2 \theta \,,
\end{equation}
where
$\theta$ is the solar neutrino mixing angle. Then the
survival probability of solar electron neutrinos is written as \cite{concha}
\begin{eqnarray}
P_{\nu_e \to \nu_e} & = &  
P^S_{e1} P^E_{1e} + P^S_{e2} P^E_{2e} + 
\nonumber \\ &&
2\, \sqrt{P^S_{e1} P^E_{1e} P^S_{e2} P^E_{2e}} 
\cos(\delta_m + \Delta m^2 L_\mathrm{vac}/2E) \,.
\label{Psun}
\end{eqnarray}
In this formula, $L_\mathrm{vac} \simeq 150 \times 10^6$ km 
is the distance between the surface
of the sun and the surface of the earth along the neutrino trajectory,
i.e. $\Delta m^2 L_\mathrm{vac}/2E$ is the phase the neutrino acquires
in vacuum. The phase acquired in matter is denoted by $\delta_m$.

Eq.~(\ref{Psun}) is completely general. One obvious questions arises: 
Under which conditions can the interference term in $P_{\nu_e \to \nu_e}$
be dropped? To investigate this point one can make a
rough estimate. The size of the neutrino production region is
approximately the diameter $\ell_\mathrm{core}$ of the solar core. 
Therefore, for $\Delta m^2 \ell_\mathrm{core}/ 2E \gtrsim 2\pi$ with
$\ell_\mathrm{core} \sim 10^5$ km, we obtain the condition 
\begin{equation}\label{dropinterference}
\Delta m^2/E > 10^{-14} \; \mathrm{eV} \quad \Rightarrow \quad 
\left\langle \cos(\delta_m + \Delta m^2 L_\mathrm{vac}/2E)
\right\rangle_\mathrm{core} = 0 \,.
\end{equation}
This argument is not
completely correct because of the matter effects, but numerical
calculations give the same result \cite{concha}. On the other hand,
experiments cannot measure the neutrino energy $E$ with infinite
precision and energy averaging in the vacuum phase occurs. If
$(\Delta m^2 L/2E) (\delta E/E) \gtrsim 2\pi$, where $\delta E$ is the
uncertainty in the measurement of $E$, with an optimistic assumption 
of $\delta E/E \sim 10^{-2}$ we again arrive at
Eq.~(\ref{dropinterference}). 

\begin{figure}[t]
\begin{center}
\includegraphics[width=8cm]{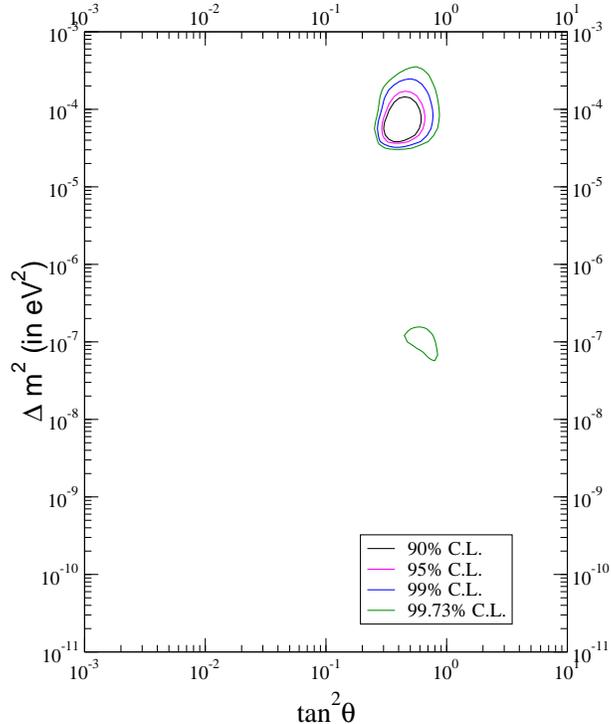}
\end{center}
\caption{Allowed regions in the $\tan^2 \theta$--$\Delta m^2$
  plane for the 2-neutrino oscillation solutions of the solar neutrino 
  deficit. The upper region represents the LMA MSW solution, the lower
  one the LOW solution. This plot showing the status after the SNO
  results but before the KamLAND result has been taken from
  Ref.~\cite{goswami}. \label{solutions}}
\end{figure}
For the LMA MSW oscillation solution of the solar neutrino puzzle with
$\Delta m^2 \sim 10^{-5}$ eV$^2$, condition
(\ref{dropinterference}) is very well fulfilled, and for the LOW
solution---named after its ``low'' mass-squared difference of 
$\Delta m^2 \sim 10^{-7}$ eV$^2$---this condition is still
fulfilled \cite{concha}. For the allowed regions in the 
$\tan^2 \theta$--$\Delta m^2$ plane of these solutions see
Fig.~\ref{solutions}. Note that this figure is relevant for the
situation after the SNO results but before the KamLAND result which
has ruled out the LOW region in the plot.
For the recent history of solar neutrino oscillations see
Ref.~\cite{drexlin}. 
For the LMA MSW solution the resonance condition (\ref{res}) plays an
important role, as we will see in the next subsection. 

\subsection{Adiabatic neutrino evolution in matter}

Now we want to discuss adiabatic neutrino evolution in matter. We
focus on solar neutrinos and confine ourselves to 2-neutrino
flavours. Solutions of the 2-neutrino differential equation
\begin{equation}\label{2-nueq}
i \frac{d}{dx} \left( \begin{array}{c} a_e \\ a_x \end{array} \right)
= H_\nu^\mathrm{mat} 
\left( \begin{array}{c} a_e \\ a_x \end{array} \right)
\end{equation}
can be found by numerical integration and, for instance, the plots
of allowed $\tan^2 \theta$--$\Delta m^2$ regions of solar
neutrino oscillations like Fig.~\ref{solutions} are usually obtained
via such numerical solutions, taking into account averaging over the
solar neutrino production region. Eq.~(\ref{2-nueq}) refers to 
$\nu_e \to \nu_x$ transitions. In the next subsection, 
in the context of three neutrino flavours, we will discuss to which 
$\nu_\mu$--$\nu_\tau$ flavour combination the amplitude $a_x$ refers.
The Hamiltonian $H_\nu^\mathrm{mat}$ is given by Eq.~(\ref{H2}).

In general, solutions of Eq.~(\ref{2-nueq}) will be
non-adiabatic. However, since the result of the KamLAND experiment we
know that the solar neutrino puzzle is solved by neutrino oscillations
corresponding to the LMW MSW solution; we will argue here that this 
solution behaves very well adiabatically. 

For the consideration of adiabaticity (see, for instance,
Ref.~\cite{schiff}) we need the
eigenvectors of $H^{\mathrm{mat}}_\nu$, which are defined 
$\forall\, x$ by 
\begin{equation}\label{eigeneq}
H^{\mathrm{mat}}_\nu(x) \psi_{mj}(x) = E_j(x) \psi_{mj}(x) \,,
\end{equation}
where 
\begin{equation}\label{eigenv}
\psi_{m1} = \left( \begin{array}{r} \cos \theta_m \\ -\sin
\theta_m \end{array} \right), \;
\psi_{m2} = \left( \begin{array}{r} \sin \theta_m \\ 
\cos \theta_m \end{array} \right).
\end{equation}
In the 2-flavour case discussed here and with the real Hamiltonian
(\ref{H2}), one parameter, the \emph{matter angle} $\theta_m$, is
sufficient to parameterize the two eigenvectors (\ref{eigenv}). 
With Eq.~(\ref{H2}) the matter angle is expressed as 
\begin{equation}\label{matterangle}
\tan 2\theta_m(x) = \frac{\tan 2\theta}{\displaystyle 1 - 
\frac{A(x)}{\Delta m^2 \cos 2\theta}} 
\quad \mathrm{with} \quad 
A(x) = 2\sqrt{2} E N_e(x) \,.
\end{equation}
Note that for $A \to 0$, i.e. negligible matter effects, we obtain 
$\theta_m \to \theta$, i.e., the matter angle becomes identical with
the (vacuum) mixing angle of $U$ of Eq.~(\ref{U2}). 

The full solution of the differential equation (\ref{2-nueq})
can be written as
\begin{equation}\label{full}
\psi(x) = \sum_{j=1,2} a_j(x) \psi_{mj}(x) e^{-i\varphi_j(x)}
\quad \mathrm{with} \quad 
\varphi_j = \int_{x_0}^x dx' E_j(x') \,.
\end{equation}
Having in mind solar neutrinos, we identify $\Delta m^2$ with the
solar mass-squared difference and use the 
initial condition that at $x_0$ a $\nu_e$ is produced. This
leads to 
\begin{equation}\label{initial}
\psi(x_0) = \left( \begin{array}{c} 1 \\ 0 \end{array} \right) 
\; \Rightarrow \left\{ \begin{array}{c}
a_1(x_0) = \cos \theta_m(x_0) \,, \\
a_2(x_0) = \sin \theta_m(x_0) \,.
                       \end{array}
\right.
\end{equation}

Eq.~(\ref{full}) is completely general. A solution is called
adiabatic if the following condition is fulfilled. 
\begin{equation}
\mathrm{Adiabaticity:} \quad a_{1,2}(x) \simeq \mathrm{constant}.
\end{equation}
In that case we have 
$a_{1,2}(x) \simeq a_{1,2}(x_0)$ and with Eq.~(\ref{initial}) we
obtain 
\begin{equation}\label{PS}
P^S_{e1} = \cos^2 \theta_m(x_0) \,, \quad 
P^S_{e2} = \sin^2 \theta_m(x_0) \,.
\end{equation}
Using now that the interference term in Eq.~(\ref{Psun}) can be dropped
for the LMA MSW solution and taking into account Eq.~(\ref{day}), we compute
\begin{eqnarray}
P_{\nu_e\to\nu_e} & = &
\cos^2 \theta_m(x_0) \cos^2 \theta +
\sin^2 \theta_m(x_0) \sin^2 \theta \nonumber  \\
& = & 
\frac{1}{2} \Big(1 + \cos 2\theta_m(x_0) \cos 2\theta \Big).
\label{Padiab}
\end{eqnarray}
This is the well-known survival probability for adiabatic neutrino
evolution from matter to vacuum. Since we used Eq.~(\ref{day}), it
does not include earth matter effects.

Let us now consider effects of 
non-adiabaticity. Plugging the full solution (\ref{full}) into
Eq.~(\ref{2-nueq}), we derive an equation for the coefficients $a_{1,2}$:
\begin{equation}\label{a12}
\frac{\mathrm{d}}{\mathrm{d}x} 
\left( \begin{array}{c} a_1 \\ a_2 \end{array} \right) =
\left( \begin{array}{cc} 0 & -\frac{\mathrm{d}\theta_m}{\mathrm{d}x}
e^{i\varphi} \\
\frac{\mathrm{d} \theta_m}{\mathrm{d}x} 
e^{-i\varphi} & 0 
\end{array} \right)
\left( \begin{array}{c} a_1 \\ a_2 \end{array} \right),
\end{equation}
where we have used the definition
\begin{equation}\label{defs} 
\varphi = \varphi_1 - \varphi_2 = 
\int_{x_0}^x dx' \Delta E 
\quad \mathrm{with} \quad 
\Delta E = E_1 - E_2 \,.
\end{equation}
With the Hamiltonian (\ref{H2}) we calculate 
\begin{equation}\label{deltaE}
\Delta E = \frac{1}{2E}\, 
\sqrt{(A-\Delta m^2 \cos 2 \theta)^2 + (\Delta m^2 \sin 2\theta)^2} \,.
\end{equation}
If along the neutrino trajectory the phase factor $\exp (i\varphi)$ in
Eq.~(\ref{a12}) oscillates very often while the change in
$\theta_m$ is of order one, then $\dot{a}_{1,2}$ will be approximately
zero. This suggest the introduction of the ``adiabaticity parameter''
\cite{haxton,parke,kuo} 
\begin{equation}\label{adiabpar}
\gamma(x) \equiv \Delta E/2|\dot{\theta}_m| \,.
\end{equation}
Therefore, we conclude
\begin{equation}
\mathrm{Adiabaticity} \; \Leftrightarrow \;\gamma \gg 1
\end{equation}
along the neutrino trajectory. In vacuum, $\gamma = \infty$, in
agreement with $a_{1,2}$ being exactly constant.
Non-adiabaticity is quantified by the probability $P_c$ for crossing from 
$\psi_{m1}$ to $\psi_{m2}$. It corrects Eq.~(\ref{Padiab}) to \cite{parke}
\begin{equation}\label{Pnonadiab}
P_{\nu_e\to\nu_e} = 
\frac{1}{2}  + \left( \frac{1}{2} - P_c \right) 
\cos 2\theta_m(x_0) \cos 2\theta \,.
\end{equation}
Again, as in Eq.~(\ref{Padiab}), this form of the survival probability
holds where Eq.~(\ref{dropinterference}) is valid. The adiabaticity
parameter $\gamma$ can be used to find a mathematically exact upper
bound on $P_c$ \cite{BGG99}.

Reverting to solar neutrinos, we use that the electron density in the
sun fulfills 
\begin{equation}\label{Ne}
N_e(x) \propto \exp{(-x/r_0)} \,,
\end{equation}
except for the inner
part of the core and toward the edge \cite{bahcall}, with 
$r_0 \simeq R/10.54 \simeq 6.6 \times 10^4 \: \mathrm{km} \leftrightarrow 3.3
\times 10^{20} \: \mathrm{MeV}^{-1}$; $R$ is the solar radius.
We want to estimate $\gamma$ for the case that the neutrino goes
through a resonance (\ref{res}) in a region where the exponential form
of the electron density provides a good approximation. Then
adiabaticity will be rather well fulfilled if it is fulfilled at the
resonance point \cite{haxton,parke,kuo,mikheyev}. There, $\gamma$ is given by 
\begin{equation}\label{adiabparres}
\gamma_\mathrm{res} = \frac{\Delta m^2 \sin^2 2\theta}%
{2 E \cos 2\theta (|\dot{A}|/A)_\mathrm{res}} \simeq
\frac{\Delta m^2 r_0}{2E} \times \cos 2\theta \tan^2 2\theta \,.
\end{equation}
For large $\theta$ and close to the solar edge this estimate for
adiabaticity is not correct (see Ref.~\cite{friedland}). 

Now we turn to the 
characteristics of the LMA MSW solar neutrino solution with 
$\Delta m^2 \sim 7 \times 10^{-5}$ eV$^2$ and $\theta \sim
34^\circ$ \cite{smirnov,goswami-review}.\footnote{Before the result of
the KamLAND experiment the mass-squared difference was a little
lower at 
$\Delta m^2 \sim 5 \times 10^{-5}$ eV$^2$---see, e.g.,
Refs.~\cite{concha,goswami}.} Writing the oscillation
length (\ref{Losc}) as 
\begin{equation}\label{LLMA}
L_\mathrm{osc} = 35 \: \mbox{km} 
\left( \frac{7 \times 10^{-5} \: \mathrm{eV}^2}{\Delta m^2} \right) 
\left( \frac{E}{1 \: \mathrm{MeV}} \right),
\end{equation}
it is evident that in this case the flavour transition occurs inside
the sun. Furthermore, with Eq.~(\ref{adiabparres}) 
we estimate $\gamma_\mathrm{res} \sim 10^3$ and the LMA MSW
oscillation solution is clearly in the adiabatic regime.
Other interesting features of the LMA MSW solution emerge by
considering the limit of increasing neutrino energy:
\begin{equation}\label{increaseE}
\renewcommand{\arraystretch}{1.2}
E \uparrow \; \Rightarrow \left\{ 
\begin{array}{l}
x_\mathrm{res}\: \mbox{moves outward}, \\
\theta_m(x_0) \to \pi/2 \,, \\
P^S_{e1} \to 0, \; P^S_{e2} \to 1 \,, \\
P_{\nu_e\to\nu_e} \to \frac{1}{2} \left( 1-\cos 2\theta \right) =
\sin^2\theta \,.
\end{array}
\right.
\end{equation}
The first line follows from Eq.~(\ref{res}) and the fact that $N_e(x)$
is monotonically decreasing from the center to the edge 
of the sun.\footnote{Note that a neutrino can go twice through a
  resonance if it is produced in that half of the sun which looks away
  from the earth.} The other properties in Eq.~(\ref{increaseE})  
are read off from Eqs.~(\ref{matterangle}), (\ref{PS}) and
(\ref{Padiab}), respectively. 
We remind the reader that the latter limit does not include earth
matter effects, therefore, it holds only during the day, just as
Eq.~(\ref{Padiab}). 
Note that from $P^S_{e2} \to 1$ we conclude that for large enough
solar neutrino energies we have a transition
\begin{equation}
| \nu_e \rangle_\mathrm{core} \to | \nu_2 \rangle_\mathrm{edge} \,.
\end{equation}

\begin{figure}[t]
\begin{center}
\includegraphics[width=11cm]{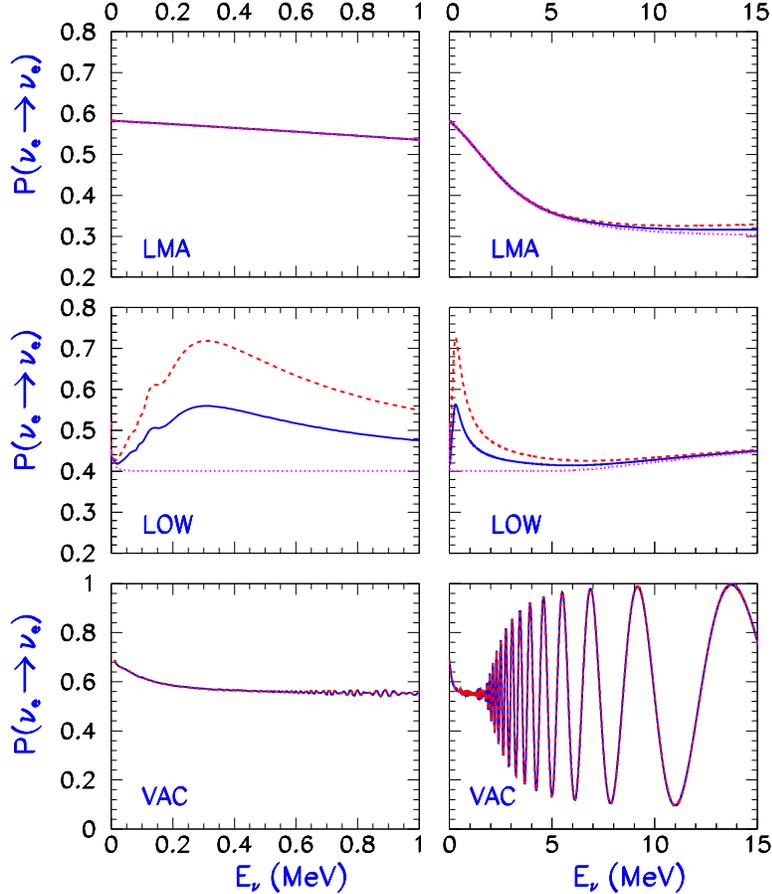}
\end{center}
\caption{Comparison of the survival probabilities for LMA MSW, LOW
  and vacuum solar neutrino oscillations for an electron neutrino
  created in the center of the sun. The figure has been taken from
  Ref.~\cite{bahcall02}. The dotted lines refer to the daytime,
  whereas the dashed line includes $\nu_e$ regeneration during the
  night, when earth matter is effective. The full line refers to the
  average survival probability. \label{survivalP}}
\end{figure}
In Fig.~\ref{survivalP} the solar $\nu_e$ survival probability is
depicted. For historical reasons, $P_{\nu_e \to \nu_e}$ is depicted not
only for the LMA MSW, but also the LOW and vacuum
oscillation\footnote{Here the oscillation length is of the order of
  the distance between sun and earth.}  
probabilities which are both ruled out now. The dotted lines refer to
daytime, i.e., when earth matter has no effect. 
The LMA MSW best fit of Ref.~\cite{bahcall02} is given by 
$\Delta m^2 = 5.0 \times 10^{-5}$ eV$^2$ and  $\tan^2 \theta = 0.42$
(after the SNO but before the KamLAND result).
Let us now make a small numerical exercise. From the best fit we get 
$\sin^2 \theta = 0.30$, which should give $P_{\nu_e \to \nu_e}$ for
large $E$. Indeed, looking at the right end of the dotted curve in the
right LMA panel where $E = 15$ MeV, we find excellent agreement. We see that
neutrino energies around 10 MeV are already ``large'' in the sense of
the limit (\ref{increaseE}). On the other hand, let us take the limit
$E \to 0$. In that limit the neutrino does not go through a resonance
because Eq.~(\ref{res}) cannot be fulfilled. Therefore, because of the
short oscillation length (\ref{LLMA}), we expect an averaged survival
probability $P_{\nu_e \to \nu_e} = 1 - \frac{1}{2} \sin^2 2\theta$ (see
Eqs.~(\ref{2tran}) and (\ref{2surv})).\footnote{We thank D.P. Roy for
a discussion on this point.} Numerically, we obtain 
$1 - \frac{1}{2} \sin^2 2\theta = 0.58$, in excellent agreement
with the left ends of the LMA panels. We have thus demonstrated here
that qualitatively the LMA MSW solution can be quite easily understood.

We have not discussed earth matter effects here. Looking at the right 
LMA panel in Fig.~\ref{survivalP}, 
we see that, at energies $E \gtrsim 10$ MeV,
during the night when earth matter
effects are operative the $\nu_e$ survival probability is a little
larger than during the day. This introduces a small day--night
asymmetry due to $\nu_e$ regeneration in earth matter. For a
qualitative understanding of this effect see Ref.~\cite{chiang}.

\paragraph*{Small $\Delta m^2$ and the limit of strong
  non-adiabaticity:} Here we want to discuss the question 
\textit{How do solar neutrinos approach vacuum oscillations 
in the limit of small $\Delta m^2$\,?}
This is a non-trivial question because for 
small $\Delta m^2$ there is always a point where
the resonance condition (\ref{res}) holds. Of course, this is a purely
academic question since we know that solar neutrinos follow the LMA
MSW solution for which the resonance does play a prominent role.
It is, however, an interesting question from the point of view of
neutrino evolution in matter in general.

When we speak of vacuum oscillations of solar neutrinos we mean
mass-squared differences of the order of 
$\Delta m^2 \sim 10^{-10}$ eV$^2$ for which the oscillation length is of
the order of the distance between sun and earth. From the resonance condition
(\ref{res}) we read off that in this case 
the resonance is very close to the solar
edge (in this context, see $N_e(x)$ in Ref.~\cite{bahcall}). 
From Eq.~(\ref{matterangle}) it follows that 
$\theta_m(x_0) = \pi/2$ and that $\dot \theta_m = 0$ holds until very
close to the resonance. It is easy to check
that the width of the resonance, i.e. the distance where the change in
$A(x)$ is of the order of $\Delta m^2 \cos 2\theta$, is roughly $r_0$,
where the electron density $N(x)$ behaves like Eq.~(\ref{Ne}). Close
to the edge of the sun the electron density drops steeper and the
resonance width is smaller. Then one can check that while the neutrino
crosses the resonance and also between the resonance and the solar
edge one can approximate the phase 
$\varphi$ of Eq.~(\ref{defs}) by a constant, which can be taken as
its value at the solar edge. This is just the
opposite of adiabaticity---see the discussion after Eq.~(\ref{a12}). 
For constant $\varphi$ the solution of Eq.~(\ref{a12}) is given by 
\begin{equation}\label{smallDm2}
\setlength{\arraycolsep}{2pt}
\left( \begin{array}{c} a_1(x) \\ a_2(x) \end{array} \right) =
\left( \begin{array}{cc} 
\cos \Delta \theta_m & - e^{i\varphi} \sin \Delta \theta_m \\
e^{-i\varphi} \sin \Delta \theta_m &  \cos \Delta \theta_m 
\end{array} \right)
\left( \begin{array}{c} a_1(x_0) \\ a_2(x_0) \end{array} \right)
\end{equation}
with 
$\Delta \theta_m \equiv \theta_m(x)-\theta_m(x_0)$. 
In the case under discussion, the solution (\ref{smallDm2}) is valid
from just before the resonance till the solar edge and in the
beginning of the vacuum. 
The initial conditions are $\theta_m(x_0) = \pi/2$ and 
$a_1(x_0) = 0$, $a_2(x_0) = 1$ . Then with Eq.~(\ref{smallDm2}), 
at the solar edge, 
we get $a_1 = e^{i\varphi} \cos \theta$, $a_2 = \sin \theta$.
Plugging this result into the full solution (\ref{full}), we find what
we expect: After the resonance, i.e. from the the solar edge onward, 
neutrinos perform ordinary vacuum oscillations; even
the matter effects accumulated in the phases $\varphi_{1,2}$ cancel. Inside
the sun before the resonance, oscillations are completely suppressed
because of the matter effects, due to $A \gg \Delta m^2 |\cos 2\theta|$.
For further details see Ref.~\cite{friedland}.

\subsection{3-neutrino oscillations and the mixing matrix}
\label{3nuoscmix}

\paragraph*{Neutrino mass spectra:} Up to now all possible neutrino mass
spectra, i.e. hierarchical, inverted hierarchy, degenerate, etc., 
are compatible with present data. By convention we have stipulated 
$m_1 < m_2$ or $\Delta m^2_{21} \equiv \Delta m^2_\odot > 0$, where we
have used the definition $\Delta m^2_{jk} = m_j^2 - m_k^2$.
Thus there are two possibilities for $m_3$ according to 
$\Delta m^2_{31}\, \raisebox{-1mm}{$\stackrel{>}{\scriptstyle <}$}\,
0$. Correspondingly, there are the two types of spectra depicted in
Fig.~\ref{spectra}: 
\begin{figure}[t]
\begin{center}
\setlength{\unitlength}{1cm}
\begin{picture}(4,4)
\put(1,1){\vector(0,1){3}}
\put(3,1){\vector(0,1){3}}
\thicklines
\put(0.7,1.3){\line(1,0){0.6}}
\put(0.7,1.5){\line(1,0){0.6}}
\put(0.7,3.5){\line(1,0){0.6}}
\put(1.1,1.1){\makebox(0,0)[l]{$m_1$}}
\put(1.1,3.65){\makebox(0,0)[l]{$m_3$}}
\put(1.1,1.65){\makebox(0,0)[l]{$m_2$}}
\put(1,0.8){\makebox(0,0)[t]{\tt normal}}
\put(0.6,1.4){\makebox(0,0)[r]{$\Delta m^2_\odot$}}
\put(0.8,2.5){\makebox(0,0)[r]{$\Delta m^2_\mathrm{atm}$}}
\put(2.7,1.3){\line(1,0){0.6}}
\put(2.7,3.3){\line(1,0){0.6}}
\put(2.7,3.5){\line(1,0){0.6}}
\put(3.1,1.1){\makebox(0,0)[l]{$m_3$}}
\put(3.1,3.65){\makebox(0,0)[l]{$m_2$}}
\put(3.1,3.1){\makebox(0,0)[l]{$m_1$}}
\put(3,0.8){\makebox(0,0)[t]{\tt inverted}}
\put(3.7,3.4){\makebox(0,0)[l]{$\Delta m^2_\odot$}}
\put(3.7,2.3){\makebox(0,0)[l]{$\Delta m^2_\mathrm{atm}$}}
\put(2,0.4){\makebox(0,0)[t]{\tt spectrum}}
\end{picture}
\end{center}
\caption{The normal and the inverted 3-neutrino mass spectrum. \label{spectra}}
\end{figure}
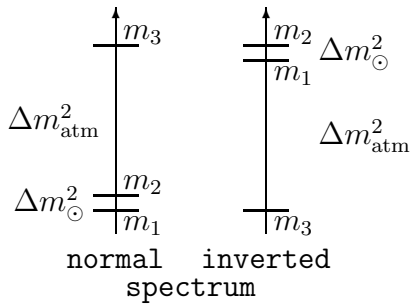
The ``normal'' and the ``inverted'' spectrum \cite{grifols}. 
The hierarchical spectrum emerges in the limit $m_1 \to 0$ of the
\texttt{normal} spectrum, whereas the 
``inverted'' hierarchy is obtained from the \texttt{inverted} spectrum
with $m_3 \to 0$.

In the 3-neutrino case the atmospheric mass-squared difference is not
uniquely defined. If by convention we use for 
$\Delta m^2_\mathrm{atm}$ the largest possible mass-squared
difference, then for the \texttt{normal} spectrum we obtain 
$\Delta m^2_\mathrm{atm} = \Delta m^2_{31}$ and for the
\texttt{inverted} case we have  
$\Delta m^2_\mathrm{atm} = \Delta m^2_{23}$.

\paragraph*{The mixing matrix:}

The most popular parameterization of the neutrino mixing matrix is
given by 
\cite{PDG} 
\begin{equation}\label{U123}
U = U_{23} U_{13} U_{12}
\end{equation}
with 
\begin{eqnarray}
U_{23} & = &
\left( \begin{array}{ccc} 1 & 0 & 0 \\
0 & c_{23} & s_{23} \\ 
0 & -s_{23} & c_{23} \end{array} \right),
\label{U23} \\
U_{13} & = & 
\left( \begin{array}{ccc} c_{13} & 0 & s_{13} e^{-i\delta} \\
0 & 1 & 0 \\ -s_{13} e^{i\delta} & 0 & c_{13} \end{array} \right),
\label{U13} \\
U_{12} & = &
\left( \begin{array}{ccc} c_{12} & s_{12} & 0 \\
-s_{12} & c_{12} & 0 \\ 0 & 0 & 1 \end{array} \right).
\label{U12}
\end{eqnarray}
The mixing angle $\theta_{12} \equiv \theta_\odot$ 
is probed in the solar neutrino
experiments and in the KamLAND experiment. The mixing angle
$\theta_{23} \equiv \theta_\mathrm{atm}$ is the relevant angle 
in atmospheric and LBL neutrino oscillation experiments. 
Up to now now effects of a non-zero angle
$\theta_{13}$ have not been seen.

\paragraph*{3-neutrino oscillations:}
The principle of 3-neutrino oscillations is a consequence of 
$\Delta m^2_\odot \ll \Delta m^2_\mathrm{atm}$ and can be summarized in the
following way. \\
\begin{center}
\fbox{\parbox{8.5cm}{
$\Delta m^2_\mathrm{atm}L/(2E) \sim 1$ $\Rightarrow$ 
$\Delta m^2_\odot L/(2E) \ll 1$\\ Solar $\nu$ oscillations frozen in
atm./LBL $\nu$ osc.!}}\\[1mm]
\fbox{\parbox{8.5cm}{
$\Delta m^2_\odot L/(2E) \sim 1$ $\Rightarrow$ 
$\Delta m^2_\mathrm{atm}L/(2E) \gg 1$\\
Atm. $\nu$ oscillations averaged in solar $\nu$ osc.!}}
\end{center}

\vspace*{2mm}

Since, at present, $\theta_{13} = 0$ is compatible with all available
neutrino data, only an upper bound on this angle can be
extracted. The two most important results in this
context are:
\begin{itemize}
\item
The non-observation of $\nu_e$ disappearance in the CHOOZ \cite{CHOOZ}
and Palo Verde \cite{paloverde} experiments; 
\item
The non-observation of atmospheric $\nu_e$ or $\bar\nu_e$
disappearance \cite{drexlin}. 
\end{itemize}
Also solar neutrino data have an effect on $\theta_{13}$, although
small, because in the 3-neutrino case the solar survival
probability is given by
\begin{equation}\label{Psolar}
P_{\nu_e\to\nu_e}^\odot \simeq 
c_{13}^4\, P^{(2)}_{\nu_e\to\nu_e}(\Delta m^2_{21}, \theta_{12}, 
c_{13}^2 N_e) + s_{13}^4 \,,
\end{equation}
where $P^{(2)}_{\nu_e\to\nu_e}$ is a 2-neutrino probability, 
calculated with the electron density $c_{13}^2 N_e$ instead of
$N_e$. The result of fits to the neutrino data is \cite{fogli}
\begin{equation}
\sin^2 \theta_{13} < 0.05 \; \mathrm{at} \; 3\sigma \,.
\end{equation}
As evident from the parameterization (\ref{U123}), CP violation
disappears from neutrino oscillations for $\theta_{13} \to 0$.

In atmospheric neutrino experiments the mixing parameter which is
extracted is 
\begin{equation}
\sin^2 2\theta_{23} \simeq 
4|U_{\mu 3}|^2 (1-|U_{\mu 3}|^2) \,,
\end{equation}
where we have taken into account the smallness of $\theta_{13}$. The
Super-Kamiokan\-de experiment finds a result \cite{SK} compatible with maximal
mixing: 
\begin{equation}
\sin^2 2\theta_{23} > 0.92 \; \mathrm{at} \; \mbox{90\% CL.}
\end{equation}
This leads to $0.60 < |U_{\mu 3}| < 0.80$. Note that 
$\theta_{23} = 45^\circ$ corresponds to 
$|U_{\mu 3}| = 1/\sqrt{2} \simeq 0.707$. 

Now we want to discuss the states into which solar and atmospheric
neutrinos are transformed. In the limit $\theta_{13} \to 0$ 
or $U_{13} \to \mathbbm{1}$ this discussion is most transparent. 
In that limit, in terms of neutrino states, atmospheric neutrino
oscillations are given by  
$| \nu_\mu \rangle \to | \nu_\tau \rangle$, 
$| \bar\nu_\mu \rangle \to | \bar\nu_\tau \rangle$, while $\nu_e$
and $\bar\nu_e$ do not oscillate because their oscillation amplitude
is given by $\sin^2 2\theta_{13}$. 
In the context of Eq.~(\ref{2-nueq}) the question was raised to
which neutrino state the amplitude $a_x$ belongs. For the
evolution of neutrino states in the sun, 
the mass eigenstate 
$|\nu_3 \rangle = s_{23} |\nu_\mu \rangle + c_{23} |\nu_\tau \rangle$
is an approximate eigenstate of $H_\nu^\mathrm{mat}$ because 
$|\Delta m^2_{31}| \gg A$ (see Eq.~(\ref{strength})). 
Therefore, the initial state $| \nu_e \rangle_\odot$ must transform 
with probability $P_{\nu_e\to\nu_e}^\odot$ into the
state orthogonal to $|\nu_3 \rangle$, from where it follows that
(without earth regeneration effects)
\begin{equation}
|\nu_e \rangle_\odot \to 
-c_{23} |\nu_\mu \rangle + s_{23} |\nu_\tau \rangle \,.
\end{equation}
Note that 
Eq.~(\ref{Psolar}) has been derived by using the approximate
eigenvector property of $|\nu_3 \rangle$, but 
corrections for non-zero $\theta_{13}$ have been made.

The primary goal of 
LBL neutrino oscillation experiments is the check of atmospheric
neutrino results, observation of $\nu_\tau$
appearance and the oscillatory behaviour in $L/E$ \cite{drexlin}.
The demonstration of the latter property would give us the final proof that
atmospheric $\nu_\mu$ and $\bar\nu_\mu$ disappearance is really an effect of
oscillations. Let us make a list what we would wish to gain from LBL
(K2K with $L = 250$ km, MINOS and CERN-Gran Sasso with 
$L = 730$ km) and very LBL experiments:
\begin{itemize}
\item[*]
Precision measurements of $\Delta m^2_\mathrm{atm}$ and $\theta_{23}$;
\item[*]
Measurement of $\theta_{13}$;
\item[*]
Verification of matter effects and distinction
between normal and inverted mass spectra;
\item[*]
Effects of $\Delta m^2_\odot$ and CP violation.
\end{itemize}
We want to stress the difficulty of measuring CP violation in neutrino
oscillations: Such an effect is only present if 
$\sin\theta_{13} \neq 0$ and if the experiment is sensitive to both
mass-squared differences, atmospheric and solar, thus the measurements must
be accurate to such a degree that they invalidate the principle of
3-neutrino oscillations mentioned in the beginning of this subsection.
Furthermore, the earth matter background in (very) LBL
experiments fakes effects of CP violation which have to be separated
from true CP violation.

Very LBL experiments are planned with super neutrino beams \cite{nakaya} and
neutrino factories \cite{geer}. Super neutrino beams are conventional
but very-high-intensity and low-energy ($E \sim 1$ GeV) beams, with
the neutrino detector slightly off the beam axis; the JHF--Kamioka
Neutrino Project with $L = 295$ km is scheduled to start in 2007
\cite{nakaya}. 
Neutrino factories are muon storage rings with straight sections from
where well-defined neutrino beams from muon decay emerge. According to
the list above, measuring $\nu_e \to \nu_\mu$ and 
$\bar\nu_e \to \bar\nu_\mu$ transitions is of particular interest; note
that $\theta_{13}$  is the only angle in the neutrino mixing matrix
which has not been measured up to now. One can easily convince oneself
that this task is tackled by measuring the so-called ``wrong-sign''
muons; e.g., 
$\mu^-_\mathrm{storage} \to \mu^+_\mathrm{detector}$ 
corresponds to $\bar\nu_e \to \bar\nu_\mu$. Neutrino factories are
under investigation. Optimization conditions for running such a
machine are considered in the ranges 
$20 \lesssim E_\mu \lesssim 50$ GeV and $L \gtrsim 3000$ km. 
For further references see, e.g., \cite{barger,freund,huber,lindner}
and citations therein.

\section{Dirac versus Majorana neutrinos}
\label{DvsM}

In this section we focus, in particular, on Majorana neutrinos. Apart
from the vanishing electric charge, also most of the popular extensions
of the SM suggest that neutrinos have Majorana nature. Majorana
particles have several interesting features, which make them quite
different from Dirac particles. Unfortunately, in practice, due to the
smallness of the masses of the light neutrinos, it is quite difficult to
distinguish between both natures: Neutrinoless $\beta\beta$-decay seems
to be the only promising road so far. In any case, neutrino
oscillations do not distinguish between Dirac and Majorana neutrinos, 
since there only the states with negative helicity enter, where this
distinction is irrelevant. Though family lepton numbers must be
violated for transitions $\nu_\alpha \to \nu_\beta$, the total
lepton number remains conserved.

\subsection{Free fields}

With two independent chiral 4-spinor fields $\psi_{L,R}$ one can
construct the usual 
\emph{Lorentz-invariant} bilinear for Dirac fields which is called mass
term:
\begin{equation}\label{Dmass}
\mbox{Dirac:} \quad 
-m \left( \bar\psi_R \psi_L + \mbox{H.c.} \right) = -m \bar\psi \psi
\quad \mbox{with} \quad \psi = \psi_L + \psi_R \,.
\end{equation}
With only one chiral 4-spinor $\psi_L$ one obtains nevertheless a 
\emph{Lorentz-invariant} bilinear with the help of the
charge-conjugation matrix $C$:
\begin{equation}\label{Mmass}
\mbox{Majorana:} \;\:
\frac{1}{2} m \psi_L^T C^{-1} \psi_L + \mbox{H.c.} = 
-\frac{1}{2} m \bar\psi \psi 
\; \mbox{with} \;
\psi = \psi_L + (\psi_L)^c.
\end{equation}
Note that from $\psi_L$ one obtains a right-handed field with the
charge-conjuga\-tion operation 
$(\psi_L)^c \equiv C \gamma_0^T \psi_L^*$; in
contrast to the Dirac case, this right-handed field is \emph{not}
independent of the left-handed field. 

The equation for free Dirac or Majorana fields in terms of the above
defined spinors $\psi$ is the same:
\begin{equation}\label{free}
\mbox{Dirac and Majorana:} \quad 
(i\gamma^\rho \partial_\rho - m) \psi = 0 \,.
\end{equation}
In the formalism used here the Majorana nature is hidden in the
\begin{equation}\label{majcond}
\mbox{Majorana condition:} \quad \psi = \psi^c \,.
\end{equation}
Thus the solution of Eq.~(\ref{free}) is found in the same way for
both fermion types. However, for Majorana neutrinos the condition
(\ref{majcond}) is imposed on the solution. This leads to the
following observations.\\
\emph{Dirac fermions:} The field $\psi$ contains 
annihilation and creation operators $a$, $b^\dagger$, respectively,
with independent operators $a$, $b$, and thus a Dirac fermion field has
particles and antiparticles with positive and negative helicities. \\
\emph{Majorana fermions:} Because of the condition (\ref{majcond}) the
annihilation and creation operators are $a$, $a^\dagger$,
respectively, and we have only particles with positive and negative
helicities. 

The mass terms (\ref{Dmass}) and (\ref{Mmass}) are written in the way
as they appear in the Lagrangian. The field equation (\ref{free}) is
obtained from the Lagrangian by variation with respect to the
independent fields. In this procedure the factor 1/2 in the Majorana mass
term is canceled by the factor of 2 which occurs in the variation of
the mass term because the fields to the left and to the right of
$C^{-1}$ are identical (see the first term in Eq.~(\ref{Mmass})). 

Up to now we have discussed the case of one neutrino with mass 
$m > 0$. Let us now consider $n$ neutrinos. Then in the Dirac case we
have the following mass term:
\begin{equation}\label{nDmass}
\mbox{Dirac:} \quad  
- \left( \bar\nu_R \mathcal{M} \nu_L + \mbox{H.c.} \right) =
- \bar\nu' \hat m \nu' \,.
\end{equation}
Here, $\mathcal{M}$ is an \emph{arbitrary} complex $n \times n$
matrix and the fields $\nu_{L,R}$ are vectors containing $n$ 
4-spinors. In order to arrive at the diagonal and positive matrix $\hat m$
we use the following theorem concerning bidiagonalization in linear
algebra. 
\newtheorem{theorem}{Theorem}
\begin{theorem}
If $\mathcal{M}$ is an arbitrary complex $n \times n$ matrix, then there
exist unitary matrices $U_{L,R}$ with 
$U_R^\dagger \mathcal{M} U_L = \hat m$ diagonal and positive.
\end{theorem}
Applying this theorem we obtain the 
\begin{equation}
\mbox{physical Dirac fields} \quad \nu' = \nu'_L + \nu'_R 
\quad \mathrm{with} \quad \nu_{L,R} = U_{L,R}\, \nu'_{L,R} \,.
\end{equation}
Note that the $U(1)$ invariance of the mass term under 
$\nu_{L,R} \to e^{i\alpha} \nu_{L,R}$ corresponds to total
lepton number conservation, provided the rest of the Lagrangian
respects this symmetry as well.

Switching to the Majorana case we have the following mass term:
\begin{eqnarray}
\mbox{Majorana:} &&
\frac{1}{2} \nu_L^T C^{-1} \mathcal{M} \nu_L + \mbox{H.c.} = 
\frac{1}{2} {\nu'}_L^T C^{-1} \hat m\, \nu'_L + \mbox{H.c.} =
\nonumber \\ && 
-\frac{1}{2} \bar\nu' \hat m \nu' \,. \label{nMmass}
\end{eqnarray}
Now the mass matrix $\mathcal{M}$ is a complex matrix which fulfills
\begin{equation}
\mathcal{M}^T = \mathcal{M} \,.
\end{equation}
This follows from the anticommutation property of the fermionic fields
and $C^T = -C$. The diagonalization of the mass term proceeds now via
a theorem of I. Schur \cite{schur}.
\begin{theorem}
If $\mathcal{M}$ is a complex symmetric $n \times n$ matrix, then
there exists a unitary matrix $U_L$ with 
$U_L^T \mathcal{M} U_L = \hat m$ diagonal and positive. 
\end{theorem}
With this theorem we obtain the 
\begin{equation}
\mbox{physical Majorana fields} \quad \nu' = \nu'_L + (\nu'_L)^c 
\quad \mbox{with} \quad \nu_L = U_L\, \nu'_L \,.
\end{equation}
The mass term (\ref{nMmass}) not only violates individual lepton
family numbers just as the Dirac mass term (\ref{nDmass}), but it also
violates the total lepton number $L = \sum_\alpha L_\alpha$.

\subsection{Majorana neutrinos and CP}

It is easy to check that 
the Majorana mass term corresponding to the mass eigenfields, the
second expression in Eq.~(\ref{nMmass}) (now we drop the primes on the
fields $\nu_L$), is invariant under \cite{BP87,wolfenstein81}
\begin{equation}\label{MCP}
\nu_{jL} \to \rho_j\, i\, C \nu_{jL}^* 
\quad \mbox{with} \quad \rho_j^2 = 1 \,.
\end{equation}
Thus Majorana neutrinos have imaginary CP parities. One can
check that $\nu_j = \nu_{jL} + (\nu_{jL})^c$ transforms into 
$\rho_j\, i\, C \nu_j^*$ under the transformation (\ref{MCP}). 
If we supplement this CP transformation by
\begin{eqnarray}
W^+_\mu & \to & -\epsilon(\mu) W^-_\mu 
\quad \mbox{with} \quad 
\epsilon(\mu) = (-1)^{\delta_{0\mu}+1} \,, \\
\ell_{\alpha L} & \to & -C \ell_{\alpha L}^* \qquad (\alpha = e, \mu,
\tau) 
\end{eqnarray} 
and require that the 
CC interactions are CP-invariant then we are lead to 
\begin{equation}\label{CPUcond}
U_M = i\, U_M^* \hat \rho 
\quad \mbox{with} \quad 
\hat \rho = \mbox{diag}\, (\rho_1,\rho_2,\rho_3) \,.
\end{equation}
We denote the mixing matrix for Majorana neutrinos by $U_M$. 
From Eq.~(\ref{CPUcond}) we derive 
\begin{equation}\label{CPCUM}
\mbox{CP invariance} \; \Rightarrow \;
U_M = O\, e^{i\frac{\pi}{4} \hat \rho} \,,
\end{equation}
where $O$ is a real orthogonal matrix. In the CP-conserving case the
matrix $O$ is relevant for neutrino oscillations since the phases in
$U_M$ of Eq.~(\ref{CPCUM}) drop out due to the rephasing invariance of
the oscillation probability (see Section \ref{neuoscvac}).
Note that one can extract an overall factor 
$\exp (i\pi/4)$ or $\exp (-i\pi/4)$ from the phase
matrix in $U_M$ and absorb it into the charged lepton fields. In this
way, one obtains another commonly used form of $U_M$ in the case of CP
conservation.  

For CP non-invariance one obtains
\begin{equation}\label{UM}
U_M = U e^{i\hat \alpha} \,.
\end{equation}
The \emph{Majorana phases} $\alpha_j$ cannot be removed by absorption into
$\nu_{jL}$ because the Majorana mass term is not invariant under such
a rephasing. However, one of the three phases can be absorbed into a
the charged lepton fields, thus only two of the Majorana phases are
physical. In neutrino oscillations only $U$ (\ref{U123}) with the 
KM phase $\delta$ is relevant. The two 
Majorana phases play an important role in the discussion of 
neutrinoless $\beta\beta$-decay and can be related to the phases which
appear in 
leptogenesis and the baryon asymmetry from leptogenesis
\cite{fukugita} (for reviews see, for instance, 
Refs.~\cite{leptogenesis-reviews}), 
though in general the low-energy CP phases are independent of 
the phases relevant in leptogenesis \cite{casas,branco} (see also
Ref.~\cite{pascoli03} and citations therein).

\subsection{Neutrinoless $\beta\beta$-decay}

Up to now, the lepton number-conserving $\beta\beta$-decay 
$(Z,A) \to (Z+2,A) + 2e^- + 2\bar\nu_e$ has been observed in direct
experiments with seven nuclides \cite{tretyak}. There is also an intensive
search for the neutrinoless $\beta\beta$ or $(\beta\beta)_{0\nu}$
decay $(Z,A) \to (Z+2,A) + 2e^-$, where the total
lepton number $L = L_e + L_\mu + L_\tau$ is violated 
($\Delta L = 2$). The distinct signal for such a decay is given by the
two electrons each with energy $E_e = (M(Z,A) - M(Z+2,A))/2$,
i.e. half of the mass difference between mother and daughter nuclide.
In contrast to $\beta\beta$-decay with neutrinos, there is no
unequivocal experimental demonstration 
for such a decay. The most stringent limit comes
from $^{76}$Ge with $T^{0\nu}_{1/2} > 1.9 \times 10^{25}$ yr 
(90\% CL) \cite{Heidelberg-Moscow}. For reviews on
$(\beta\beta)_{0\nu}$ decay experiments see Refs.~\cite{tretyak,vogel}.

There are several mechanisms on the quark level which induce two
transitions $n \to p$ in a nucleus without emission of neutrinos. 
Here we discuss some of the popular ones (see Ref.~\cite{mohapatra}
for a review of $(\beta\beta)_{0\nu}$ decay mechanisms).
\begin{itemize}
\item 
\texttt{Effective Majorana neutrino mass:} This mechanism uses the
Majorana nature of neutrinos and can schematically be depicted in the
following way:
$$
\begin{array}{c}
d \to u + e^- + \bar\nu_e \\
d \to u + e^- + \bar\nu_e
\end{array}
\raisebox{-2pt}{$\big]$} \quad \raisebox{-2pt}{Wick contraction}
$$
This Wick contraction can be performed because $\nu_j^c = \nu_j$ and,
therefore, the electron-neutrino field can be written as
\begin{equation}
\nu_{eL} = \sum_j (U_M)_{ej} \nu_{jL} =  
\sum_j (U_M)_{ej} \frac{\mathbbm{1} - \gamma_5}{2} C (\bar\nu_j)^T \,.
\end{equation}
Thus one actually contracts $\nu_j$ with $\bar\nu_j$ $\forall j$. In
other words, the relevant neutrino propagator is given by the
expression
\begin{eqnarray}
&& 
\langle 0 | T \nu_{eL}(x_1) \nu_{eL}^T(x_2) | 0 \rangle =  
\nonumber \\
&&
-\sum_j (U_M)_{ej}^2 m_j\, i \int \frac{d^4p}{(2\pi)^2} 
\frac{e^{-ip \cdot (x_1-x_2)}}{p^2 - m_j^2}\, 
\frac{\mathbbm{1}-\gamma_5}{2}\, C \,.
\end{eqnarray}
One can show that it is allowed to neglect $m_j^2$ in the denominator
of the $\nu_j$ propagators \cite{vogel}, which leads then to the 
\emph{effective Majorana neutrino mass}
\begin{equation}\label{effMmass}
\langle m \rangle \equiv 
\sum_j \left( U_M \right)_{ej}^2 m_j \,.
\end{equation}
For this mechanism of $(\beta\beta)_{0\nu}$ decay, the decay amplitude
is proportional to $|\langle m \rangle|$. 

The present lower bounds on $T^{0\nu}_{1/2}$ for $^{76}$Ge lead to an
upper on $\left| \langle m \rangle \right|$ varying between 0.33 and
1.35 eV, depending on the models used to calculate the nuclear matrix
element \cite{Heidelberg-Moscow,IGEX}.
Future experiments want to probe $|\langle m \rangle|$ down to 
a few$\,\times\, 0.01$ eV \cite{vogel}.

In the case of CP conservation (see Eq.~(\ref{CPCUM}))
the effective Majorana neutrino mass is given by 
$\langle m \rangle = i\sum_j O_{ej}^2\, \rho_j m_j$. For opposite 
signs of $\rho_j = \pm1$, i.e. 
opposite CP parities, cancellations in $\langle m \rangle$ 
naturally occur.
\item
\texttt{Intermediate doubly charged scalar:} In left-right symmetric
models, doubly charged scalars occur in the scalar gauge triplets and
induce $(\beta\beta)_{0\nu}$ decay via the mechanism \cite{mohapatra}
depicted symbolically in the following way:
$$
\begin{array}{c}
d \to u + S^-\, \raisebox{-3pt}{$\searrow$} \\[1.5mm]
d \to u + S^-\, \raisebox{+3pt}{$\nearrow$}
\end{array}
H^{--} \to e^- e^-
$$
\item
\texttt{SUSY with R-parity violation:} Within this framework, several
mechanisms for inducing $(\beta\beta)_{0\nu}$ decay exist
\cite{mohapatra}. The gluino ($\tilde g$) exchange mechanism uses the
Majorana nature of gauginos \cite{hirsch}: 
$$
\renewcommand{\arraystretch}{0.7}
\begin{array}{rll}
d \to e^- + \tilde u & & \\
                     & \!\!\!\!\raisebox{+4pt}{$\searrow$}\: u +
                     \tilde g & 
\\[-1mm]
                     & \!\!\!\!\raisebox{-4pt}{$\nearrow$}\: u + \tilde g & 
\raisebox{10pt}{$\Big]$} \: 
\raisebox{8pt}{Wick contraction} \\
d \to e^- + \tilde u & &
\end{array}
$$
In this picture, 
$\tilde u$ is an up squark. 
\end{itemize}
For the check of the Majorana nature of neutrinos, $(\beta\beta)_{0\nu}$
decay is most realistic possibility. Another line which is pursued is the
search for $\bar\nu_e$ from the sun \cite{smy}. 
Both types of experiments aim at finding $|\Delta L| = 2$ processes.

In view of several mechanisms for $(\beta\beta)_{0\nu}$ decay where
some even do not require neutrinos, the question arises if an
experimental confirmation of this decay really signals Majorana nature
of neutrinos. The question was answered affirmatively by 
Schechter and Valle \cite{schechter}, and Takasugi
\cite{takasugi}. Their statement is the following:
\textit{If neutrinoless $\beta\beta$-decay exists, then neutrinos have
  Majorana nature, irrespective of the mechanism for
  $\beta\beta$-decay.} We follow the argument of Takasugi and show the
following: \textit{If $(\beta\beta)_{0\nu}$ decay exists, then a 
Majorana neutrino mass term cannot be forbidden by a symmetry.} \\
\textbf{Proof:} We assume for simplicity that there is no neutrino
mixing. As will be seen the arguments can easily be extended to the
mixing case. We introduce phase factors $\eta$ and assume the symmetry 
\begin{equation}\label{symmm}
\begin{array}{cc}
u_L \to \eta_u u_L\,, & d_L \to \eta_d d_L\,, \\
e_L \to \eta_e e_L\,, & \nu_{eL} \to \eta_\nu \nu_{eL} \,,
\end{array}
W^+ \to \eta_W W^+ \,. 
\end{equation}
Invariance of the weak interactions under this symmetry give the
following relations:
\begin{eqnarray}
\bar\nu_{eL} \gamma^\rho e_L W^+_\rho & \Rightarrow &
\eta_\nu^* \eta_e \eta_W = 1 \,, \label{wl} \\
\bar u_L \gamma^\rho d_L W^+_\rho & \Rightarrow & 
\eta_u^* \eta_d \eta_W = 1 \,.   \label{wq}
\end{eqnarray}
Now we assume the existence of $(\beta\beta)_{0\nu}$ decay, which
introduces a further relation:
\begin{eqnarray}\label{bbdecay}
d_L + d_L \to u_L + u_L + e_L + e_L & \Rightarrow &
(\eta_d^* \eta_u \eta_e)^2 = 1
\end{eqnarray}
Eqs.~(\ref{wl}) and  (\ref{wq}) together lead to 
$\eta_\nu = \eta_d^* \eta_u \eta_e$. 
Thus with Eq.~(\ref{bbdecay}) we arrive at $\eta_\nu^2 = 1$, and the mass
term $\propto \nu_{eL}^T C^{-1} \nu_{eL}$ 
cannot be forbidden by $\nu_{eL} \to \eta_\nu \nu_{eL}$.
Q.E.D.

The above argument does not necessarily show that a Majorana neutrino
mass term must appear, but in general this will be the case for 
renormalizeability reasons: All terms of dimension 4 or less which are
compatible with the symmetries of the theory have to be included in the
Lagrangian.

\subsection{Absolute neutrino masses}

With neutrino oscillations, the lightest neutrino mass which we denote
by $m_0$ cannot be determined. Nowadays three sources of
information on $m_0$ are used: 
$^3$H decay, $(\beta\beta)_{0\nu}$ decay and cosmology \cite{weiler}. 
For an extensive review see Ref.~\cite{grifols}.

\paragraph{Tritium and neutrinoless $\beta\beta$-decays:}
In the decay $^3\mbox{H} \to\, ^3\mbox{He} + e^- + \bar\nu_e$, with an 
energy release $E_0 \simeq 18.6$ keV, the region near the endpoint of
the electron recoil energy spectrum is investigated. Deviations from
the ordinary $\beta$ spectrum with a massless $\bar\nu_e$ are
searched for. Assuming that the $\bar\nu_e$ has a mass $m_\beta$, the
Mainz and Troitsk experiments have both obtained an upper limit
$m_\beta < 2.2$ eV at 95\% CL. 
The future KATRIN experiment plans to have a sensitivity 
$m_\beta \gtrsim 0.35$ eV \cite{drexlin}. 

Since we know that the electron neutrino field is a linear combination
of neutrino mass eigenfields because of neutrino mixing (see
Eq.~(\ref{mixing})), the $^3$H decay rate is actually a sum over the
decay rates of the massive neutrinos multiplied by $|U_{ej}|^2$. 
Thus the question arises for meaning of $m_\beta$ which is extracted
from the data. Farzan and Smirnov \cite{farzan} have argued in the
following way. Firstly, the bulk of data from which the bound on $m_\beta$
is derived does \emph{not} come from the very end of recoil spectrum. 
Secondly, the energy resolution in $^3$H experiments, even for the
KATRIN experiment, is significantly coarser than 
$\sqrt{\Delta m^2_\mathrm{atm}}$.
Therefore, the three masses $m_j$ can effectively be replaced by one 
mass $m_\beta$ and the different expressions for $m_\beta$ found in
the literature, e.g. $m_\beta \equiv \sum_j |U_{ej}|^2 m_j$, are
all equivalent. The simplest choice is $m_\beta = m_0$ (for other
papers on this subject see citations in \cite{farzan}).
Thus the results of the $^3$H decay experiments 
give information on the lightest neutrino mass $m_0$.

In order to make contact between $m_0$ and $(\beta\beta)_{0\nu}$
decay, the following assumptions are made: Clearly, one has to
assume Majorana neutrino nature, but also that the 
dominant mechanism for $(\beta\beta)_{0\nu}$ decay proceeds via the
effective Majorana neutrino mass 
$\left| \langle m \rangle \right|$ (\ref{effMmass}),
which is then expressed as a
function of $m_0$ and the parameters of the Majorana neutrino mixing
matrix $U_M$ (\ref{UM}). 
Note that this mechanism \textit{must} be present for Majorana
neutrinos, it might only be that it is not the dominant one.
For the \texttt{normal} spectrum we have $m_0 = m_1$ and 
$m_0 = m_3$ for the \texttt{inverted} spectrum (see Fig.~\ref{spectra}).
\begin{eqnarray}
\mbox{Normal spectrum:} && \nonumber \\
|\langle m \rangle| & = & \left| \left( m_0\, c_{12}^2 + 
\sqrt{m_0^2 + \Delta m^2_\odot}\, s_{12}^2\, e^{i\beta_1} \right) c_{13}^2 +
\right. \nonumber \\[1mm] && \left.
\sqrt{m_0^2 + \Delta m^2_\mathrm{atm}}\, s_{13}^2\, e^{i\beta_2} \right|;
\label{eff-normal} \\
\mbox{Inverted spectrum:} && \nonumber \\
|\langle m \rangle| & = & \left| \left( \sqrt{m_0^2 + 
\Delta m^2_\mathrm{atm} - \Delta m^2_\odot}\, 
c_{12}^2\, e^{i\beta_1} + \right. \right.
\nonumber \\[1mm] && \left. \left.
\sqrt{m_0^2 + \Delta m^2_\mathrm{atm}}\, s_{12}^2\, e^{i\beta_2}
\right) c_{13}^2 + 
m_0\, s_{13}^2 \right|.
\label{eff-inverted}
\end{eqnarray}
In these expressions, $\beta_1$ and $\beta_2$ are general CP-violating
phases which arise according to the form of $U_M$.\footnote{They are
simple functions of the Majorana phases and of $\delta$.} For
numerical estimates it is useful to have in mind that 
$\sqrt{\Delta m^2_\mathrm{atm}} \sim 0.05$ eV and 
$\sqrt{\Delta m^2_\odot} \sim 0.008$ eV. For references see, for
instance, \cite{pascoli02,petcov} and citations therein.

\begin{figure}[t]
\begin{center}
\epsfig{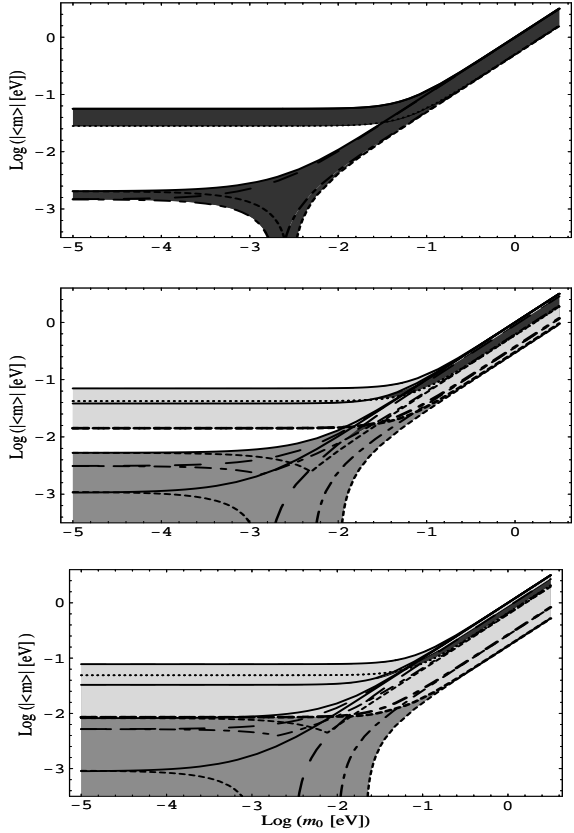}
\end{center}
\caption{Effective Majorana neutrino mass $| \langle m \rangle|$ versus
  lightest mass $m_0$. The figure is taken from
  Ref.~\cite{pascoli02}. In the upper panel best fit values for the
  neutrino oscillation parameters have been used and only the phases $\beta_1$
  and $\beta_2$ have been varied. (A non-zero best fit value for
  $s_{13}$ is used in this plot.) In the lower two panels the uncertainty in 
  the determination of the oscillation parameters has been taken into
  account. For the details and for the precise meaning of the lines in
  the shaded regions, which 
  correspond to CP-conserving choices of the phases $\beta_{1,2}$, see
  Ref.~\cite{pascoli02}. \label{m0-meff}}
\end{figure}
In order to plot $|\langle m \rangle|$ against $m_0$, 
the values of neutrino oscillation parameters are used as input.
The phases $\beta_{1,2}$ are free parameters; for 
CP conservation the possible values of $\beta_{1,2}$ are zero or
$\pi$. The plot of Fig.~\ref{m0-meff} has been taken from
Ref.~\cite{pascoli02}. In the upper panel, best fit values of the
neutrino oscillation parameters are used, whereas in the lower
two panels the uncertainty in these parameters is taken into
account. The lower band on the left sides of these panels correspond
to the normal spectrum, whereas the upper band to the inverted
spectrum. 

What does this figure teach us? One can use experimental upper 
bounds on $|\langle m \rangle|$ and $m_0$ and compare with the allowed
regions in the plot. 
It has been put forward in Ref.~\cite{czakon} that such a comparison,
even if $(\beta\beta)_{0\nu}$ decay is not found experimentally, 
might reveal the neutrino nature:
If the minimum of $|\langle m \rangle|$ predicted from $m_\beta$ and
the oscillation parameters exceeds the experimental upper bound on
$|\langle m \rangle|$, then the neutrino must have 
Dirac nature. There are, however, several difficulties with such a
procedure: Other mechanisms for
$(\beta\beta)_{0\nu}$ decay could destroy the validity of such a
comparison; $m_\beta$ will not reach the non-degenerate neutrino
mass region even with the KATRIN experiment, and the same difficulty 
applies to the region with 
$|\langle m \rangle|$ below 0.01 eV; moreover, imprecise determination
of the oscillation parameters blur the picture as seen from the lower
two panels of Fig.~\ref{m0-meff}. However, the chances for this
approach would be good, if the KATRIN experiment
would find a non-zero $m_\beta$. On the other hand, assuming the
validity of the Majorana hypothesis, Fig.~\ref{m0-meff} might 
give us information about the type of neutrino mass spectrum if
$|\langle m \rangle|$ is measured or a sufficiently stringent
experimental bound on it is derived \cite{pascoli02}.

\paragraph*{Neutrino masses and cosmology:} 
A very interesting bound on the sum over all neutrino masses is
provided by the large-scale structure of the universe \cite{colless} and the
temperature fluctuations of the cosmic microwave background (CMB)
\cite{spergel}. Usually, energy densities $\rho_i$ in the 
universe are given as fractions 
$\Omega_i \equiv \rho_i/\rho_\mathrm{cr}$ of the critical energy
density $\rho_\mathrm{cr}$ of the universe. Today's energy density of
non-relativistic neutrinos and antineutrinos is given by 
(see, for instance, \cite{raffelt})
\begin{equation}\label{numass-cosmos}
\Omega_\nu h^2 = \sum_j m_j/(93.5\: \mathrm{eV}) \,,
\end{equation}
where $h \simeq 0.7$ is the Hubble constant in units of 
100 km\,s$^{-1}$\,Mpc$^{-1}$. 
Of the three active neutrinos, at least two of them are
non-relativistic today since 
$\sqrt{\Delta m^2_\odot} \sim 0.008$ eV and
$\sqrt{\Delta m^2_\mathrm{atm}} \gg \sqrt{\Delta m^2_\odot}$,
while the neutrino temperature today corresponds to 
$1.7 \times 10^{-4}$ eV \cite{raffelt}.

Roughly speaking, an upper bound on $\Omega_\nu$ is obtained in the
following way. Hot dark matter tends to erase primordial density
fluctuations. The theory of structure formation together with data on
large-scale structures (distribution of galaxies and galaxy clusters), 
mainly from the 2 degree field Galaxy Redshift Survey \cite{colless}, 
gives an upper bound on $\Omega_\nu/\Omega_m$, where $\Omega_m$ is the
total matter density. On the other hand, data on the 
temperature fluctuations of the CMB can
be evaluated with the so-called $\Lambda$CDM model (a flat
universe with cold dark matter and a cosmological constant) and allows
in this way to determine a host of quantities like $h$, $\Omega_m$,
$\Omega_b$ (the baryon density), etc. Since the quantities extracted
in this way agree rather well with determinations based on different 
methods and assumptions \cite{spergel}, 
the $\Lambda$CDM model is emerging as the
standard model of cosmology \cite{sarkar}.  
Evaluation of the recent CMB results of the Wilkinson Microwave
Anisotropy Probe (WMAP) and of the large-scale structure data 
give the impressive bound \cite{spergel} 
$\Omega_\nu h^2 < 0.0076$ (95\% CL); 
with Eq.~(\ref{numass-cosmos}) this bound is
reformulated in terms of neutrino masses:
\begin{equation}\label{mumassbound}
\sum_j m_j < 0.7 \; \mathrm{eV}\;  (95\% \mathrm{CL}).
\end{equation}
For a single neutrino we have thus $m_j < 0.23$ eV, 
since neutrinos with masses in the range of a few tenths eV have to be
degenerate. 

Note that the $\Lambda$CDM model gives the result 
$\Omega_m \sim 0.3$ and $\Omega_\Lambda \sim 0.7$; the
latter quantity is the energy density associated with the
cosmological constant $\Lambda$. Furthermore, one finds  
$\Omega_\nu \lesssim 0.015$ and $\Omega_b \sim 0.04 \div 0.05$, thus
the main contribution to $\Omega_m$ must consist of hitherto undetermined
dark matter components.

\subsection{Neutrino electromagnetic moments}

The effective Hamiltonians for neutrinos with magnetic moments and
electric dipole moments (electromagnetic moments) are given by 
the following expressions:
\begin{eqnarray}
\mbox{Dirac:} &&
\mathcal{H}_{\mathrm{em}}^D =
\frac{1}{2} \bar{\nu}_R \lambda\,
\sigma^{\mu \nu} \nu_L F_{\mu \nu} + \mathrm{H.c.}, 
\label{diracEM} \\
\mbox{Majorana:} &&
\mathcal{H}_{\mathrm{em}}^M = 
- \frac{1}{4} \nu_L^T C^{-1} \lambda\,
\sigma^{\mu \nu} \nu_L F_{\mu \nu} + \mathrm{H.c.} 
\label{majoranaEM}
\end{eqnarray}
In the Majorana case, from
$(C^{-1} \sigma^{\mu\nu})^T =  C^{-1} \sigma^{\mu\nu}$ and the 
anticommutation property of $\nu_L$, it follows that  
$\lambda^T = -\lambda$ and only transition moments are allowed. 
In the Dirac case, the electromagnetic moment matrix $\lambda$ is an
arbitrary $3 \times 3$ matrix. Usually, a decomposition of 
$\lambda$ is made into $\lambda = \mu - id$ with 
$\mu^\dagger = \mu$ being the MM matrix and 
$d^\dagger = d$ the EDM matrix.
However, since the neutrino 
mass eigenstates are experimentally not accessible, the 
distinction MM/EDM is completely unphysical in terrestrial experiments. This
holds also largely for solar neutrinos, though in the evolution of the
solar neutrino state the squares of the neutrino masses enter, and one can
shift phases from the mixing matrix to the electromagnetic moment
matrix and vice versa, which makes the distinction MM/EDM
phase-convention-dependent. For a thorough discussion of this point see
Ref.~\cite{GS00}. 

Bounds on $\lambda$ from solar and reactor neutrinos are obtained via 
elastic $\nu e^-$ scattering where, due to the helicity flip in the
Hamiltonians (\ref{diracEM}) and (\ref{majoranaEM}), the cross section
is the sum of the weak and electromagnetic cross sections \cite{bardin}:
\begin{equation}\label{w+em}
\frac{d \sigma}{dT} = \frac{d \sigma_w}{dT} + 
\frac{d \sigma_\mathrm{em}}{dT} 
\quad \mbox{with} \quad
\frac{d \sigma_\mathrm{em}}{dT} = 
\frac{\alpha^2 \pi}{m_e^2 \mu_B^2} 
\left( \frac{1}{T} - \frac{1}{E} \right) \mu_\mathrm{eff}^2 \,.
\end{equation}
In the most general form, the effective MM is given by
\cite{GS00} 
\begin{equation}\label{effMM}
\mu_\mathrm{eff}^2 = 
a_-^\dagger \lambda^\dagger \lambda a_- + 
a_+^\dagger \lambda \lambda^\dagger a_+ \,.
\end{equation}
In Eq.~(\ref{w+em}), 
$T$ is the kinetic energy of the recoil electron, $\mu_B$ is the Bohr
magneton, and
the flavour 3-vectors $a_\pm$ describe the neutrino helicity states at
the detector. The expression (\ref{effMM}) is basis-independent, thus
it does not matter in which basis, flavour or mass basis, the
quantities $\lambda$, $a_\pm$ and the Hamiltonians (\ref{diracEM}) and
(\ref{majoranaEM}) are perceived \cite{GS00}. 

In the following we will consider reactor and solar
neutrinos. If the detector is close to the 
reactor, one simply has (in the flavour basis) 
$a_-=0$, $a_+ = (1,0,0)^T$, as the reactor
emits antineutrinos with positive helicity. For 
solar neutrinos, the effective MM will, in general, depend on the
neutrino energy $E$. 
Bounds on the neutrino electromagnetic moments have been obtained from
both reactor \cite{derbin} and solar neutrinos \cite{beacom,mohanty}.

From now on, we concentrate on Majorana neutrinos, where the
antisymmetric matrix $\lambda$ contains only three complex parameters
(in the Dirac case there are nine parameters). Thus, $\lambda$ can
be written as
\begin{equation}\label{lambda}
\lambda_{\alpha\beta} = 
\varepsilon_{\alpha\beta\gamma} \Lambda_\gamma \,,
\end{equation}
in either basis, flavour or mass basis. For the approximations used in the
calculation of the effective MM for solar neutrinos and the LMA MSW
solution, $\mu_\mathrm{LMA}^2$, we refer the reader to
Ref.~\cite{GMSTV}. The result is given by
\begin{equation}\label{effMM-LMA}
\mu_\mathrm{LMA}^2(E) = |\mathbf{\Lambda}|^2 - |\Lambda_2|^2 +
P^S_{e1} \left( |\Lambda_2|^2 - |\Lambda_1|^2 \right).
\end{equation}
The probability $P^S_{e1}$ is defined before Eq.~(\ref{day}). 
For reactor neutrinos the effective MM is immediately obtained as 
\begin{eqnarray}
\mu_\mathrm{reactor}^2 & = & 
|\lambda_{e\mu}|^2 + |\lambda_{e\tau}|^2 = 
|\Lambda_\mu|^2 + |\Lambda_\tau|^2  \nonumber \\
& = &
|\mathbf{\Lambda}|^2 -
c_{12}^2 |\Lambda_1|^2 - s_{12}^2 |\Lambda_2|^2  
-2 s_{12} c_{12} |\Lambda_1| |\Lambda_2| \cos\delta' \,,
\label{effMM-reactor}
\end{eqnarray}
where the second line is derived from the first line by using 
$ \mathbf{\Lambda} = U_M \tilde\mathbf{\Lambda}$.
The phase $\delta'$ is composed of 
$\delta' = \arg (\Lambda_1^* \Lambda_2) + \alpha_2 - \alpha_1$ with
the Majorana phases $\alpha_j$ defined in Eq.~(\ref{UM}).
Eq.~(\ref{lambda}) defines the vectors 
$\mathbf{\Lambda} = ( \Lambda_\alpha)$ and
$\tilde\mathbf{\Lambda} = (\Lambda_j)$ in the flavour and mass
basis, respectively. The length of $\mathbf{\Lambda}$ occurs in both
effective moments, Eqs.~(\ref{effMM-LMA}) and
(\ref{effMM-reactor}). Note, however, that 
\begin{equation}
|\mathbf{\Lambda}|^2 = \frac{1}{2} \mathrm{Tr}\, (\lambda^\dagger
\lambda) \quad \Rightarrow \quad 
|\mathbf{\Lambda}| = |\tilde\mathbf{\Lambda}| \,,
\end{equation}
i.e., the length of the vector of the electromagnetic moments is
basis-indepen\-dent. 

In Ref.~\cite{GMSTV} a bound on $|\mathbf{\Lambda}|$ is derived, 
i.e., \emph{all} three transition moments are bounded by using
as input the rates of the solar neutrino experiments, 
the shape of Super-Kamiokande recoil electron energy spectrum and the
results of reactor experiments at Bugey (MUNU) and Rovno \cite{derbin}. 
As statistical procedure a Bayesian method is used with flat priors and 
minimization of $\chi^2$ with respect to 
$\theta_{12}$, $\Delta m^2_\odot$, $|\Lambda_1|$, $|\Lambda_2|$, 
$\delta'$, in order to extract a probability distribution for
$|\mathbf{\Lambda}|$. The details are given in Ref.~\cite{GMSTV},
where the 90\% CL bounds
\begin{equation}\label{MMbound}
|\mathbf{\Lambda}| < \bigg\{ 
\begin{array}{cl}
6.3 \times 10^{-10} \mu_B & (\mbox{solar data})\,, \\
2.0 \times 10^{-10} \mu_B & (\mbox{solar} + \mbox{reactor data})
\end{array}
\end{equation}
are presented.

Some comments are at order \cite{GMSTV}:
\begin{itemize}
\item[$\triangleright$]
We want to stress once more that 
the bounds in Eq.~(\ref{MMbound}) apply to
$
|\mathbf{\Lambda}|^2 = 
|\lambda_{e\mu}|^2 + |\lambda_{\mu\tau}|^2 + |\lambda_{\tau e}|^2 =
\sum_{j<k} |\lambda_{jk}|^2 \,.
$
Therefore, all transition moments are bounded in a basis-independent way.
\item[$\triangleright$]
In Eq.~(\ref{effMM-LMA}), in the limit $P^S_{e1} \to 0$, the
quantity $|\Lambda_2|$ drops out of the effective MM. Therefore, for
small $P^S_{e1}$, solar neutrino data become less stringent for 
$|\Lambda_2|$ and thus for $|\mathbf{\Lambda}|$ as well. This is 
the case for Super-Kamiokande data (see the first bound in
Eq.~(\ref{MMbound})) where $P^S_{e1}$ is small due to 
the relatively high neutrino energy (see Eq.~(\ref{increaseE})). 
\item[$\triangleright$]
Reactor data give a good bound on $|\Lambda_2|$ and are, therefore,
complementary to present solar neutrino data. 
\item[$\triangleright$]
The BOREXINO experiment \cite{BOREXINO} 
could improve the second bound in Eq.~(\ref{MMbound}) 
by nearly one order of magnitude
because it is sensitive to relatively low neutrino energies where 
$P^S_{e1} \sim 0.5$.
\end{itemize}

\section{Models for neutrino masses and mixing}
\label{models}

\subsection{Introduction and scope}

For a start, we present the problems of model building
for neutrino masses and mixing by
asking appropriate questions and formulating answers, if available. 
\begin{enumerate}
\item
Can neutrino masses and mixing be \emph{accommodated} in a model? \\
This is no problem; the simplest possibility is given by the SM + 3
right-handed neutrino singlets $\nu_R$ + $L$ conservation which allows
to accommodate arbitrary masses and 
mixing of Dirac neutrinos in complete analogy to the quark sector. 
\item
Why are neutrino masses much smaller than the charged lepton masses? \\
There are two popular proposals for a solution to this problem:
\begin{itemize}
\item Radiative neutrino masses;
\item Seesaw mechanism \cite{seesaw,senjanovic}.
\end{itemize}
\item
Can one reproduce the special features of $\nu$ masses and mixing? \\
Let us list the features one would like to explain:
\begin{itemize}
\item[F1 ] $\theta_\odot \sim 34^\circ$ (large but non-maximal);
\item[F2 ] $\theta_\mathrm{atm} \simeq 45^\circ$;
\item[F3 ] $|U_{e3}|^2 \equiv s_{13}^2 \lesssim 0.05$;
\item[F4 ] $\Delta m_\odot^2/\Delta m_\mathrm{atm}^2 \sim 0.03$.
\end{itemize}
We have used the obvious notation where the subscripts $\odot$ and
atm refer to solar and atmospheric neutrinos, respectively. 
\end{enumerate}

To explain the listed features is the most difficult and largely unsolved
task. There are myriads of textures or models---for reviews see
Ref.~\cite{reviews-models}. One of the problems is that there are
still not enough clues where to start model building.
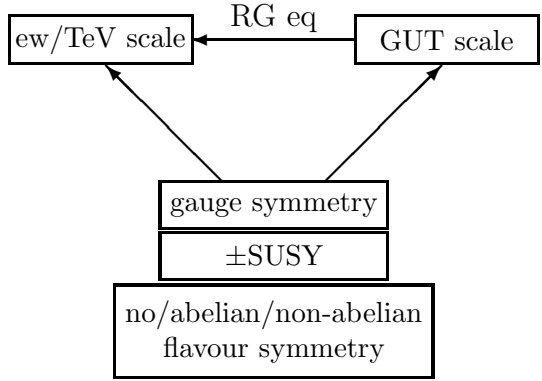
\begin{figure}[t]
\begin{center}
\setlength{\unitlength}{1cm}
\begin{picture}(7,5)
\thicklines
\put(0,4){\framebox(2.4,0.6){\small ew/TeV scale}}
\put(4.6,4){\framebox(2.4,0.6){\small GUT scale}}
\put(4.6,4.3){\vector(-1,0){2.2}}
\put(1.4,-0.18){\framebox(4.2,1.2)%
{\shortstack{\small no/abelian/non-abelian\\\small flavour symmetry}}}
\put(2,1.12){\framebox(3,0.6){\small $\pm$SUSY}}
\put(2,1.8){\framebox(3,0.6){\small gauge symmetry}}
\put(2.8,2.43){\vector(-1,1){1.53}}
\put(4.2,2.43){\vector(+1,1){1.53}}
\put(3.5,4.4){\makebox(0,0)[b]{RG eq}}
\end{picture}
\end{center}
\caption{Symbolical presentation of the variety of possibilities for
  model building. The upper two blocks show the two energy scales at which
  the symmetries, indicated in the three blocks below, could
  apply. RG eq stands for renormalization group equation, $\pm$ for
  with/without. \label{building}} 
\end{figure}
Symbolically, the difficulties of model building are presented in
Fig.~\ref{building}. If the symmetries apply at the GUT scale, the
renormalization group equation transports relations from the
GUT scale to the low scale; this transport could contribute to
generate some of the desired features---for a review see
Ref.~\cite{chankowski}. It is completely unknown if 
the explanation of the features F1-4 is independent of general fermion
mass problem or not. In the following we will assume independence. 

The scope of the following sections is to discuss some simple
extensions of the SM by addition of scalar multiplets and of right-handed
neutrino singlets $\nu_R$. All these extensions will yield 
Majorana neutrino masses. We will only discuss the lepton sector.

\subsection{The Standard Model with additional scalar multiplets}

In the SM there are only two types of lepton multiplets:
$$
\begin{array}{ccccl}
& SU(2) & \!\!\times\!\! & \! U(1) & \\
D_L & \underline{\frac{1}{2}} && Y = -1 & \mbox{left-handed doublets} \\
\ell_R & \underline{0} && Y = -2 & \mbox{right-handed singlets}
\end{array}
$$
Here $Y$ is the hypercharge and the underlined numbers indicate the
weak isospin of the multiplet.

By forming all possible leptonic bilinears one obtains the possible
scalar gauge multiplets which can couple to fermions
\cite{konetschny}: 
$$
\begin{array}{cr@{\,=\,}lccl}
\bar D_L \otimes \ell_R & 
\underline{\frac{1}{2}} \otimes \underline{0} & \underline{\frac{1}{2}} &
Y = -1 & \:\phi & \mbox{doublet} \\
D_L \otimes D_L & 
\underline{\frac{1}{2}} \otimes \underline{\frac{1}{2}} & 
\underline{0} \oplus \underline{1} & Y = -2 &
\left\{ \begin{array}{c} \eta^+ \\ \!\!\Delta \end{array} \right. &
\!\!\!\begin{array}{l} \mbox{singlet} \\ \mbox{triplet} \end{array} \\
\ell_R \otimes \ell_R & \underline{0} \otimes \underline{0} &
\underline{0} & Y = -4 & \quad k^{++} & \mbox{singlet}
\end{array}
$$
The hypercharge in this table refers to the leptonic bilinear.
In the following we will discuss extensions
of the SM with these scalar multiplets (except the trivial extension
where only Higgs doublets are added).

\paragraph*{The Zee model:}
This model \cite{zee1,zee2} is defined as 
SM with $2\phi$ + $\eta^+$ and, therefore, has the Lagrangian
\begin{equation}\label{Lazee}
\mathcal{L} = \cdots 
+ \left[ f_{\alpha\beta} D_{\alpha L}^T C^{-1} i\tau_2 D_{\beta L}\, \eta^+ -
\mu\, \phi_1^\dagger \tilde \phi_2 \eta^+ + \mbox{H.c.} \right],
\end{equation}
where the dots indicate the SM part and terms of the Higgs potential
which are not interesting for the following discussion. 
Note the antisymmetry of the coupling matrix:
$f_{\alpha\beta} = -f_{\beta\alpha}$.

Since no fermionic multiplets have been added to the SM, the
ensuing neutrino masses can only be of Majorana type. Thus the total
lepton number $L$ must be broken, otherwise neutrino masses will be
strictly forbidden. Let us assign a lepton number to $\eta^+$ from its
Yukawa couplings in Eq.~(\ref{Lazee}). Then we have the following list
of lepton numbers of the multiplets of the Zee model:
\begin{equation}\label{Lzee}
\begin{array}{c|cccc}
  & D_L & \ell_R & \phi_{1,2} & \eta^+ \\ \hline
L & 1   & 1      & 0          & -2
\end{array}
\end{equation}
Thus, $L$ is indeed explicitly violated by the term 
$\phi_1^\dagger \tilde \phi_2 \eta^+$ in the Lagrangian. Such a term
can only be formed if \emph{two} Higgs doublets are present 
(if $\phi_1 = \phi_2 = \phi$, then $\phi^\dagger \tilde \phi \equiv
0$). In the Zee model a neutrino mass matrix $\mathcal{M}_\nu$ appears at
the 1-loop level.

A lot of work has been done on the restricted Zee model
\cite{wolfenstein-zee} where only one Higgs doublet, say $\phi_1$,
couples to the leptons. In this particularly simple version one has 
$\mathcal{M}_\nu \propto \left( (m_\alpha^2 - m_\beta^2)
f_{\alpha\beta} \right)$, where $m_{\alpha,\beta}$ denote the
charged lepton masses. Note that without loss of generality we have
chosen a basis where the charged lepton mass matrix is diagonal.
The restricted Zee model is practically ruled out now because it
allows only maximal solar mixing \cite{jarlskog};
furthermore, it requires serious fine-tuning, namely
$m_\tau^2 |f_{e\tau}| \simeq m_\mu^2 |f_{e\mu}|$; 
finally, the smallness of the neutrino masses has to be achieved by 
$|f_{\alpha\beta}| \lesssim 10^{-4}$.
If both $\phi_{1,2}$ couple to the leptons, then non-maximal solar
mixing is admitted \cite{balaji}, the fine-tuning problem is somewhat
alleviated, but the third point remains. For further recent literature
on the Zee model see, e.g. Ref.~\cite{koide}

\paragraph*{The Zee-Babu model:} This model is defined as
\cite{zee2,babu} 
SM + $\eta^+$ + $k^{++}$ and has the Lagrangian
\begin{equation}\label{Lbabu}
\begin{array}{rcl}
\mathcal{L} & = & \cdots + \left[ 
f_{\alpha\beta} D_{\alpha L}^T C^{-1} i\tau_2 D_{\beta L}\, \eta^+
\right. \\[0.5mm]  && \left. +
h_{\alpha\beta} \ell_{\alpha R}^T C^{-1} \ell_{\beta R}\, k^{++} - 
\mu\, \eta^-\eta^- k^{++} + \mbox{H.c.} \right]
\end{array}
\end{equation}
with a symmetric coupling matrix 
$h_{\alpha\beta} = h_{\beta\alpha}$. In addition to the assignments
(\ref{Lzee}), we have $L(k^{++}) = -2$. Again, $L$ must be explicitly
broken, which is now achieved by the $\mu$-term in Eq.~(\ref{Lbabu}).

Since in this model there is only one Higgs doublet, $L$ is still
conserved at the 1-loop level (see previous section on the Zee model)
and the neutrino mass matrix appears at the 
2-loop level: 
$
\mathcal{M}_\nu \propto \tilde f \hat m_\ell {\tilde h}^* \hat m_\ell
\tilde f$ with 
$\tilde f = (f_{\alpha\beta})$, $\tilde h = (h_{\alpha\beta})$ 
and $\hat m_\ell = \mbox{diag}\,(m_e,m_\mu,m_\tau)$.
This model has the following properties \cite{macesanu}. Since 
$\tilde f$ is antisymmetric, the lightest neutrino mass is zero; 
the solar LMA MSW solution and, e.g., a hierarchical neutrino mass spectrum
require the fine-tuning
$|h_{\mu\mu}| : |h_{\mu\tau}| : |h_{\tau\tau}| \simeq 
1 : (m_\mu/m_\tau) : (m_\mu/m_\tau)^2$; 
all scalar masses are in the TeV range; in order to reproduce the
neutrino masses inferred from atmospheric and solar data one needs 
$|f_{\alpha\beta}|$, $|h_{\alpha\beta}| \lesssim 0.1$, thus 
neutrino masses are naturally small as a consequence of the 2-loop mechanism;
finally, rare decays like $\tau \to 3\mu$, $\mu \to e \gamma$ should
be within reach of forthcoming experiments. 

We note that from the Zee model and Zee-Babu model we have seen that
models with radiative neutrino mass generation are prone to excessive
fine-tuning because the hierarchy of the charged lepton masses works
against the features needed in the neutrino sector.

\paragraph*{The triplet model:} This model is defined by adding a
scalar triplet $\Delta$ to the SM and has the Lagrangian 
\begin{equation}\label{Ltriplet}
\begin{array}{rcl}
\mathcal{L} & = & \cdots + \left[
\frac{1}{2} g_{\alpha\beta} D_{\alpha L}^T C^{-1} i\tau_2 \Delta
D_{\beta L} + \mbox{H.c.}\right] \\  
&& - \frac{1}{2} M^2 \mbox{Tr}\, \Delta^\dagger \Delta - 
\left( \mu\, \phi^\dagger \Delta \tilde \phi + \mbox{H.c.} \right)
\end{array}
\end{equation}
The electric charge eigenfields and the VEV of the triplet are given as 
\begin{equation}
\Delta = \left( \begin{array}{cc} H^+ & \sqrt{2} H^{++} \\
\sqrt{2} H^0 & -H^+ \end{array} \right),
\quad \langle H^0 \rangle_0 = \frac{1}{\sqrt{2}} v_T \,,
\end{equation}
respectively. Note that $g_{\alpha\beta} = g_{\beta\alpha}$. As in the
previous two models, we can make the assignment 
$L(\Delta) = -2$, then $L$ is explicitly broken by the $\mu$-term in
Eq.~(\ref{Ltriplet}). The original model without the $\mu$-term and
spontaneous $L$ breaking \cite{gelmini} is ruled out by the non-discovery
of the Goldstone boson and a light scalar at LEP. 
This model leads to the tree-level neutrino
mass matrix $\mathcal{M}_\nu = v_T (g_{\alpha\beta})$. The LEP data
require $|v_T/v| \lesssim 0.03$ \cite{erler}, where $v$ is the VEV of
the SM Higgs doublet. However, if the coupling constants
$g_{\alpha\beta}$ are about $0.01 \div 0.1$, then 
in order to obtain small neutrino masses the triplet VEV must be much
smaller, namely $v_T \sim 0.1 \div 1$ eV. 
There are two ways to get a small $v_T$: Firstly, one can assume that 
$M, \: |\mu| \gg v$, then $|v_T| \simeq |\mu| v^2/M^2$ (scalar or type
II seesaw mechanism) \cite{senjanovic,ma}; secondly, with 
$M \sim v, \: |\mu| \ll v$ one has $|v_T| \sim |\mu|$
\cite{pfeiffer}. In some sense the mechanism for obtaining small
neutrino masses in the triplet model is analogous to the seesaw
mechanism \cite{seesaw,senjanovic} (see subsequent subsection) since
in both cases one has to introduce a second scale much larger
(smaller) than the electroweak scale in order to generate small
neutrino masses. 

\subsection{The seesaw mechanism}

The seesaw mechanism \cite{seesaw,senjanovic} is implemented in the
simplest way in the SM + $3\, \nu_R$ + $L$ violation. 
This is primarily an extension in the fermion sector of the SM. 
Note that one could also choose two or more than three right-handed
singlets $\nu_R$. 
The starting point is the Lagrangian 
\begin{equation}\label{Lseesaw}
\begin{array}{rcl}
\mathcal{L} & = & \cdots
- \left[ \sum_j \left( 
\bar \ell_R \phi_j^\dagger \Gamma_j + 
\bar \nu_R {\tilde\phi}_j^\dagger \Delta_j \right) D_L
+ \mbox{H.c.}\right] \\[2mm] &&
+ \left( \frac{1}{2}\, \nu_R^T C^{-1} \! M_R^* \nu_R + \mbox{H.c.} \right),
\end{array}
\end{equation}
where the mass matrix $M_R$ of the right-handed singlets must be
symmetric due to their fermionic anticommutation property.
The number $n_H$ of Higgs doublets is irrelevant for the seesaw mechanism. 
Defining the mass matrices 
\begin{equation}
M_\ell = \frac{1}{\sqrt{2}} \sum_j v_j^\ast \Gamma_j \,,
\quad 
M_D = \frac{1}{\sqrt{2}} \sum_j v_j \Delta_j \,,
\end{equation}
where $M_\ell$ is the mass of the charged leptons and $M_D$ is the
so-called ``Dirac mass matrix'' for the neutrinos, the 
total \emph{Majorana} mass matrix for all six left-handed fields
$\omega_L$ is given by \cite{66}
\begin{equation}\label{MD+M}
\setlength{\arraycolsep}{2pt}
\mathcal{M}_{D+M} = \left( \begin{array}{cc}
0 & M_D^T \\ M_D & M_R 	   \end{array} \right)
\quad \mbox{with} \quad \omega_L = 
\left( \begin{array}{c} \nu_L \\ C (\bar\nu_R)^T \end{array} \right).
\end{equation}
The VEVs $v_j$ fulfill 
$v = \sqrt{|v_1|^2 + \ldots |v_{n_H}|^2} = (\sqrt{2}\, G_F)^{-1/2} 
\simeq 246$ GeV. 

The basic assumption for implementing the seesaw mechanism is 
$m_D \ll m_R$, where $m_{D,R}$ are the scales of $M_{D,R}$, respectively. 
With this assumption,
one obtains the mass matrix of the light neutrinos
\begin{equation}\label{Mnu}
\mathcal{M}_\nu = -M_D^T M_R^{-1} M_D \,,
\end{equation}
which is valid up to corrections  of order $(m_D/m_R)^2$. The 
mass matrix of heavy neutrinos is given by 
$\mathcal{M}_\nu^\mathrm{heavy} = M_R$. Diagonalizing the mass
matrices by 
\begin{equation}
(U_R^\ell)^\dagger M_\ell U_L^\ell = \hat m_\ell \,, \quad
V^T \mathcal{M}_\nu V = \hat m \,,
\end{equation}
the neutrino mixing matrix for the light neutrinos is given by 
\begin{equation}\label{seesaw-mixing}
U_M = (U_L^\ell)^\dagger V \,.
\end{equation}
As discussed earlier, phases multiplying $U_M$ from the left are
unphysical because they can be absorbed into the charged lepton
fields. 

The seesaw mechanism contains three sources for neutrino mixing: 
$M_\ell$, $M_D$ and $M_R$. Therefore, it is a rich playground  
for model building. It can also be combined
with radiative neutrino mass generation---for an example see next
subsection. Let us discuss the order of magnitude of the scale
$m_R$. If we choose as a typical neutrino mass  
$m_\nu \sim \sqrt{\Delta m^2_\mathrm{atm}} \sim 0.05$ eV and assume 
$m_D \sim m_{\mu,\tau}$, we obtain $m_R \sim 10^8 \div 10^{11}$ GeV. On
the other hand, if $m_D$ is of the order of the electroweak scale,
then $m_R \sim 10^{15}$ GeV and there could be a possibility to
identify it with the GUT scale.

\subsection{Combining the seesaw mechanism with radiative
  mass generation}

Let us now consider the
SM with \emph{two} Higgs doublets $\phi_{1,2}$, add a \emph{single}
field $\nu_R$ and allow for $L$ violation \cite{GN89,GN00}.
We do not employ any flavour symmetry.

In this case the Yukawa coupling matrices 
$\Delta_{1,2}$ and $M_D$ are $1 \times 3$ matrices (vectors!). Here we
give a precise meaning to the  scale $m_D$ and 
identify it with the length of the vector $M_D$. $L$ violation is
induced by the Majorana mass 
term with mass $M_R \gg m_D$ of the neutrino singlet $\nu_R$. Then  
at the tree level the seesaw mechanism is operative:
\begin{equation}\label{tree}
\mbox{Tree level:} \quad 
m_1 = m_2 = 0 \,, \; m_3 \simeq m_D^2/M_R \,.
\end{equation}
That two masses are zero at the tree level is a consequence of $M_D$
being a $1 \times 3$ matrix, which, therefore, maps two vectors onto 0;
this feature is then operative in $\mathcal{M}_{D+M}$ (\ref{MD+M}) and 
$\mathcal{M}_\nu$ (\ref{Mnu}). 

At 1-loop level, $m_2$ becomes non-zero by neutral-scalar exchange,
bit $m_1$ remains zero:
\begin{equation}\label{1loop}
\mbox{1-loop:} \quad m_1 = 0 \,, \, \; 
m_2 \sim \frac{1}{16\pi^2}\, m_3\,
  \frac{M_0^2}{v^2} \ln\, (M_R/M_0)^2 \,.
\end{equation}
In this order-of-magnitude relation for $m_2$ \cite{GN00}, the mass 
$M_0 \sim v$ is a typical scalar mass (there are three
physical neutral scalars in this model). A general discussion for 
arbitrary numbers of left-handed lepton doublets, right-handed
neutrinos singlets and Higgs doublets is found in \cite{GN89}. The
full calculation of the dominant 1-loop corrections to the seesaw
mechanism is presented in \cite{GL02}.

The model has the following properties. It 
predicts a hierarchical spectrum, therefore 
$\sqrt{\Delta m^2_\odot/\Delta m^2_\mathrm{atm}} \simeq m_2/m_3 
\stackrel{\mathrm{exp.}}{\sim} 0.17$. 
It gives the correct order of magnitude of $m_2/m_3$ with 
$1/16\pi^2 \simeq 0.0063$, $M_0^2/v^2 \sim 1$, $\ln (M_R/M_0) \sim
10$. The most interesting property of these 1-loop corrections to the
seesaw mechanism is the fact that the suppression relative to the tree
level terms is given solely by the loop integral factor $1/16\pi^2$
\cite{GN89,GL02}. 
The mixing angles are undetermined, but without
fine-tuning they will be large in general. Consequently, the model has
no argument for $\theta_\mathrm{atm} \simeq 45^\circ$ and small angle
$\theta_{13}$; these must be reproduced by tuning the parameters of
the model. This is easily achieved because in good approximation 
$|U_{\alpha 3}| = |M_{D \alpha}|/m_D$ \cite{GN00}.

It is interesting to note that R-parity-violating supersymmetric
models have a built-in seesaw mechanism where the heavy Majorana
neutrinos are replaced by the neutralinos which are also Majorana
particles. See, e.g., Ref.~\cite{SUSY} and citations therein.

\subsection{A model for maximal atmospheric neutrino mixing}

Her we want to discuss a model based on tree-level seesaw masses. 
While in the previous model the emphasis was on explaining the ratio 
$\Delta m^2_\odot/\Delta m^2_\mathrm{atm}$ and we assumed to obtain
the mixing angles by tuning of model parameters, here we take the
opposite attitude. One of the problems of obtaining maximal
atmospheric neutrino mixing and large but non-maximal solar neutrino
mixing by a symmetry is that the mixing
matrix (\ref{seesaw-mixing}) has a contribution also from the charged
lepton sector. Now we introduce a framework where this problem is
avoided by having $M_R$ as the 
only source of neutrino mixing. 

\paragraph*{The framework:} We start with the Lagrangian
(\ref{Lseesaw}) and allow for an arbitrary number $n_H$ of Higgs
doublets. Then how can one avoid flavour-changing interactions via
tree-level scalar interactions? 
We assume that the family lepton numbers 
$L_{e,\mu,\tau}$ are conserved in all terms of dimension 4 in the
Lagrangian but 
\textit{$L_\alpha$ is softly broken by the $\nu_R$ mass term}.
This allows one to kill several birds with one stone: Flavour-changing
neutral interactions are forbidden at the tree level, 
$M_\ell$ and $M_D$ are diagonal
and neutrino mixing stems exclusively from $M_R$ \cite{GL01}, as
announced above. However, soft breaking of lepton numbers occurs at
the \emph{high} scale $m_R$. It has been shown in Ref.~\cite{GL02a}
that this assumption yields a perfectly viable theory with interesting
properties: All 1-loop flavour-changing vertices are finite because of
soft $L_\alpha$ breaking; for $n_H > 1$, such vertices where the boson
leg is a neutral scalar and the exchanged boson is a charged scalar do
not decouple in the limit $m_R \to \infty$ (for $n_H = 1$ there is
decoupling); diagrams with a 
$\gamma$ or a $Z$ and box diagrams always decouple. As a consequence, the 
amplitudes of $\mu \to e \gamma$, $Z \to e^- \mu^+, \ldots$ are
suppressed by $1/m_R^2$, whereas, e.g., the amplitude of $\mu \to 3e$
tends to a constant for $m_R \to \infty$ and is suppressed---though
much less than the previous amplitudes---because it contains a product
of four Yukawa couplings. The decay rate of the latter process in this
framework might be within reach of
forthcoming experiments. For details see Ref.~\cite{GL02a}. 

\paragraph*{Maximal atmospheric neutrino mixing:}
Within the framework of soft $L_\alpha$-breaking we introduce now a
$\mathbbm{Z}_2$ symmetry: 
\begin{equation}\label{Z2}
\mathbbm{Z}_2: \quad 
D_{\mu L} \leftrightarrow D_{\tau L}, \:
\nu_{\mu R} \leftrightarrow \nu_{\tau R}, \: \mu_R \leftrightarrow
\tau_R \,.
\end{equation}
This symmetry makes $M_D$ and $M_R$ $\mathbbm{Z}_2$-invariant and,
therefore, transfers to the neutrino mass matrix
$\mathcal{M}_\nu$. Thus we obtain the result  
\begin{equation}\label{Z2massmatrix}
\setlength{\arraycolsep}{3pt}
\mathcal{M}_\nu =
\left( \begin{array}{ccc}
x & y & y \\ y & z & w \\ y & w & z
\end{array} \right)
\; \Rightarrow \;
U = \left( \begin{array}{ccc}
\scriptstyle \cos \theta & \scriptstyle \sin \theta & \scriptstyle 0 \\
-\frac{\sin \theta}{\sqrt{2}} & \frac{\cos \theta}{\sqrt{2}} 
& \frac{1}{\sqrt{2}} \\
\frac{\sin \theta}{\sqrt{2}} & -\frac{\cos \theta}{\sqrt{2}} 
& \frac{1}{\sqrt{2}}
\end{array} \right),
\end{equation}
where $\theta \equiv \theta_{12}$ is the solar mixing
angle. Summarizing, we have constructed a model where the solar mixing
angle is free and without fine-tuning it will be large but non-maximal; 
atmospheric mixing is maximal; $|U_{e3}| = s_{13} = 0$. 

The above results are stable under radiative corrections because the
mass matrix (\ref{Z2massmatrix}) was realized by the symmetries
$U(1)_{L_e} \times U(1)_{L_\mu} \times U(1)_{L_\tau}$, which are softly
broken, and $\mathbbm{Z}_2$, which has to be broken spontaneously in order to
achieve $m_\mu \neq m_\tau$. This can be done with a minimum of three
Higgs doublets and an auxiliary $\mathbbm{Z}'_2$, without destroying
the form of the mass matrix (\ref{Z2massmatrix}) \cite{GL01}. Since
the $\mathbbm{Z}_2$ of Eq.~(\ref{Z2}) does not commute with
$L_{\mu,\tau}$, the full group generated by our symmetries is
non-abelian \cite{GL03}. The essential
features of this model can be embedded in an $SU(5)$ GUT \cite{GL03}.

\section{Conclusions}
\label{summary}

In recent years we have witnessed great progress in neutrino
physics. Eventually, it has been confirmed that the 
solar neutrino puzzle is solved by neutrino oscillations, 
first conceived by Bruno Pontecorvo in 1957. At the same time, matter
effects in neutrino oscillations---as occurring in 
the LMA MSW solution---have turned out to play a decisive role.
It is firmly believed that also the atmospheric neutrino
deficit problem is solved by neutrino oscillations though for the time
being the final proof is still missing but is expected to be provided
soon by LBL experiments.

Despite the great progress, which provides us with a first glimpse
beyond the SM, there are still many things 
we would like to know. For instance, Are neutrinos Dirac or Majorana
particles? What are the absolute neutrino masses? Of what type is the neutrino
mass spectrum? Do neutrinos have sizeable magnetic moments? 

As for field-theoretical models of neutrino masses and mixing, theorists are
groping in the dark. As it has turned out, neutrino mass spectra and
the neutrino mixing matrix are very different from the quark sector.
Though there are many ideas, very few of them account naturally for
some of the neutrino properties. Despite the big increase in our
knowledge about neutrinos even basic questions for model building have
no answer at the moment. Some of the basic questions are the
following: Is the neutrino mass and mixing problem independent of the
general fermion mass problem? 
Are neutrino masses small by radiative, seesaw or other mechanisms?
Is the solution for the neutrino mass and mixing problem situated at
the TeV scale or the GUT scale? Do we need a flavour symmetry for its
solution and, if yes, of what type is this symmetry?

We want to stress that it is no problem to \emph{accommodate} neutrino
masses and mixing in theories, but the problem is to \emph{explain} 
the specific features for neutrinos.
As for the neutrino nature, there is a 
theoretical bias toward Majorana nature, e.g. from the 
seesaw mechanism and GUTs, in particular, from GUTs based on
$SO(10)$. For the time being, 
there are simply not enough clues for a definite mechanism for
neutrino masses and mixing. Among others, 
possible future clues provided by experiment would be  
an atmospheric mixing angle very close to $45^\circ$, which would point
toward a non-abelian flavour symmetry; knowledge of the 
value of $|U_{e3}| = s_{13}$ and the neutrino mass spectrum, 
which would teach us more about the mass
matrix; discovery of scalars with masses $\lesssim$ 1 TeV, which would
show that fermion masses are most probably generated by the Higgs mechanism;
discovery of SUSY partners of ordinary particles, which would assure us
that we have to take SUSY into account in model building. 
Evidence for flavour-changing decays like $\mu \to e \gamma$, $\mu \to
3e$, etc.\ would also give a valuable input. 

At any rate, ongoing and future experiments will continue to provide
us with exciting results, which will enhance the prospects of
constructing viable mechanisms for explaining the specific neutrino
features. 

\vspace{2mm}

\noindent
\textbf{Acknowledgements:} I wish to thank the Organizing Committee
for the invitation to Schladming and A. Bartl, G. Ecker, M. Hirsch and
H. Neufeld for discussions.

\end{document}